\documentclass[twocolumn]{aastex7}

\usepackage{amsmath}	
\usepackage{amssymb}	
\usepackage{ulem}   
\usepackage{makecell} 
\usepackage{multirow}

\newcommand{\Hentropy}{Permutation entropy}
\newcommand{\Ccomplexity}{Statistical complexity}
\newcommand{\hentropy}{permutation entropy}
\newcommand{\ccomplexity}{statistical complexity}

\newcommand{\pecc}{PECCARY}
\newcommand{\peccDef}{Permutation Entropy and statistiCal Complexity Analysis for astRophYsics}

\newcommand{\tser}{time-series}
\newcommand{\Tser}{Time-series}

\newcommand{\tstep}{time-step}
\newcommand{\Tstep}{Time-step}

\newcommand{\edel}{sampling interval}
\newcommand{\Edel}{Sampling interval}

\newcommand{\edim}{sampling size}
\newcommand{\Edim}{Sampling size}

\newcommand{\hcplane}{$HC$-plane}

\newcommand{\edelsymbol}{\ell}

\newcommand{\tscal}{timescale}
\newcommand{\Tscal}{Timescale}
\newcommand{\tscals}{timescales}
\newcommand{\Tscals}{Timescales}
\newcommand{\pattscal}{pattern \tscal}
\newcommand{\Pattscal}{Pattern \tscal}

\newcommand{\dset}{dataset}

\newcommand{\Lyp}{Lyapanov}
\newcommand{\stocity}{stochasticity}
\newcommand{\tpat}{t_{\rm pat}}
\newcommand{\tdur}{t_{\rm dur}}
\newcommand{\cmplx}{complex}

\newcommand{\tnat}{t_{\rm nat}}
\newcommand{\tdurtnat}{\tdur/\tnat}
\newcommand{\tpattnat}{\tpat/\tnat}

\newcommand{\HMaxPer}{H_{\rm per}^{\rm max}}
\newcommand{\HMinPer}{H_{\rm per}^{\rm min}}
\newcommand{\hcval}{[H,C]}

\newcommand{\Hofl}{H(\edelsymbol)}
\newcommand{\Cofl}{C(\edelsymbol)}
\newcommand{\Hcurve}{$H$-curve}
\newcommand{\Ccurve}{$C$-curve}
\newcommand{\HandCcurves}{$H$- and \Ccurve s}

\newcommand{\henon}{H\'enon}
\newcommand{\galpy}{\texttt{galpy}}

\usepackage{xcolor}
\definecolor{dcolor}{RGB}{19,204,26}
\definecolor{kcolor}{rgb}{0.54, 0.17, 0.89}
\definecolor{scolor}{RGB}{0, 166, 222}
\definecolor{amaranth}{rgb}{0.9, 0.17, 0.31}
\definecolor{pescyColor}{RGB}{171, 0, 60}
\definecolor{black}{RGB}{0,0,0}

\renewcommand{\sout}[1]{\unskip}

\newcommand{\kdc}[1]{\unskip}
\newcommand{\dsc}[1]{\unskip}
\newcommand{\shc}[1]{\unskip}

\received{August 1, 2024}
\revised{May 5, 2025}
\accepted{May 14, 2025}
\submitjournal{ApJ}

\shorttitle{PECCARY}
\shortauthors{Hyman, Daniel, and Schaffner}

\begin{document}


\title{\pecc: A novel approach for characterizing orbital complexity, stochasticity, and regularity}

\correspondingauthor{S\'oley Hyman}

\author[0000-0002-6036-1858, gname="S\'oley", sname="Hyman"]{S\'oley \'O. Hyman}
\affiliation{Steward Observatory and Department of Astronomy, University of Arizona\\933 N. Cherry Ave., Tucson, AZ 85721, USA}
\email[show]{soleyhyman@arizona.edu}

\author[0000-0003-2594-8052, gname="Kathryne", sname="Daniel"]{Kathryne J. Daniel}
\affiliation{Steward Observatory and Department of Astronomy, University of Arizona\\933 N. Cherry Ave., Tucson, AZ 85721, USA}
\affiliation{Department of Physics, Bryn Mawr College, Bryn Mawr, PA 19010, USA}
\email{kjdaniel@arizona.edu}

\author[0000-0002-9180-6565, gname="David", sname="Schaffner"]{David A. Schaffner}
\affiliation{Department of Physics, Bryn Mawr College, Bryn Mawr, PA 19010, USA}
\email{dschaffner@brynmawr.edu}

\begin{abstract}
\peccDef\ (\pecc) is a computationally inexpensive, statistical method by which any \tser\ can be characterized as predominantly regular, complex, or stochastic. Elements of the \pecc\ method have been used in a variety of physical, biological, economic, and mathematical scenarios, but have not yet gained traction in the astrophysical community. This study introduces the \pecc\ technique with the specific aims to motivate its use in and optimize it for the analysis of astrophysical orbital systems.
\pecc\ works by decomposing a time-dependent measure, such as the $x$-coordinate or orbital angular momentum \tser, into ordinal patterns.
Due to its unique approach and statistical nature, \pecc\ is well suited for detecting preferred and forbidden patterns (a signature of chaos), even when the chaotic behavior is short-lived or when working with a relatively short-duration \tser\ or small sets of \tser\ data. 
A variety of examples are used to demonstrate the capabilities of \pecc. These include mathematical examples (sine waves, varieties of noise, well-known chaotic functions), a double pendulum system, and astrophysical tracer particle simulations with potentials of varying intricacies. Since the adopted \tscal\ used to diagnose a given \tser\ can affect the outcome, a method is presented to identify an ideal sampling scheme, constrained by the overall duration and the natural \tscal\ of the system. The accompanying \pecc\ Python package and its usage are discussed.

\end{abstract}


\section{Introduction} \label{sec:intro}
\peccDef\ (\pecc) is a statistical method used to characterize a \tser\ as regular, stochastic (i.e., random or noisy), or complex, and identify its relevant \tscals\ \citep{Bandt2002,Rosso07,Weck2015}. The use of \hentropy\ and \ccomplexity\ measures has been gaining traction in a wide variety of physical, biological, and mathematical scenarios, including plasma turbulence~\citep{maggs2013,gekelman2014,Maggs2015,Weck2015,Zhu2017}, solar wind and space plasma~\citep{suyal2012,Weck2015,Ribeiro2017,olivier:2019,Weygand:2019,Good:2020}, geological processes~\citep{Donner2015}, river flow~\citep{serinaldi2014,Thaxton2018}, economic trends~\citep{Zunino2010,Bariviera2013,araujo:2020}, and biological or medical rhythms~\citep{Jordan2008,Li2010,Aronis2018MNRAS}.

One of the drivers of the evolution of a dynamical orbital system depends on the relative fraction and distribution of regular and complex orbits, as well as the \tscals\ associated with each.
Indeed, the formulation of chaos theory itself is firmly rooted in the study of chaotic behavior in astrophysical dynamical systems \citep[e.g.,~the three-body problem,][]{Poincare1891} and continues to inform studies of the secular evolution of galaxies \citep[e.g.,][]{Fux01,Pichardo03,Patsis06,MA11,Valluri16}, planetary systems \cite[e.g.,][]{Gladman93,Malhotra93,ST93,Astakhov03,LQ11,DPH13}, and black hole dynamics \cite[e.g.,][]{Contopoulos90,SM00,Merritt04}, to name a few.
The role of chaos in the evolution of dynamical systems is not the same as that of stochastic processes \citep[whether physical or computationally induced, see][for various treatments]{Pfenniger86,KW94,MCC06,SD09,Neyrinck22}, though they can be nearly indistinguishable in practice  \citep{Rosso07} and quite often both are confusingly labeled ``stochastic." Differentiating between the two is particularly relevant in understanding the difference between dynamical processes and issues that arise from the limited resolution in discretized computation, such as shot noise \cite[e.g.,][]{SvdW00,Dehnen01,VRG03,SD09,Sellwood13}.

The \pecc\ method is able to discern the nature of fluctuations in a \tser\ through its decomposition into a distribution of the occurrence frequency of patterns (described in Section~\ref{subsec:ord-patts}). \pecc\ can be set apart from well-known methods for determining regions of orbital chaos or irregularity, such as \Lyp\ exponential divergence \citep[e.g.,][]{Pichardo03}, Kolmogorov-Arnold-Moser (KAM) theory analysis \citep{Weinberg15a,Weinberg15b}, frequency map analysis \citep[e.g.,][]{LFC92,PL96,PL98,ValluriMerritt98,Valluri12,Price-Whelan16,BeraldoeSilva19}, and surface of section analysis \citep[e.g.,][]{Martinet74,Athanassoula83} since it is optimized to identify chaotic behavior on relatively short \tscals\ and is agnostic to underlying physics. \pecc\ expands the toolbox that astrophysical researchers have at their disposal for understanding dynamical systems. It provides an analysis technique that is suitable in situations where long-standing traditional techniques may not be as applicable, such as simulations with time-dependent potentials (e.g., a slowing bar or in systems that are accreting mass).

This work introduces the theoretical framework of \pecc\ and explores its applicability and limitations in astrophysical systems. \S\ref{sec:overview} gives an overview of the theory, how ordinal patterns are determined, how the metrics of \hentropy\ and \ccomplexity\ are computed, and the usage of the \hcplane.
\S\ref{sec:usage-interpretation} discusses the usage, interpretation, and limitations of the method, as well as an idealized sampling scheme. \S\ref{sec:examples} demonstrates the capabilities of \pecc\ via a variety of mathematical, physical, and astrophysical examples. 
\S\ref{sec:future} provides an outlook into future work and tests to be done with \pecc, and \S\ref{sec:conclusion} summarizes the conclusions.

\section{Overview of the \pecc\ Method}\label{sec:overview}
\pecc\ is comprised of two different statistical measures: \hentropy\ and \ccomplexity. These two measures were developed in the early 2000s \citep[e.g.,][]{Bandt2002,Rosso07} as a way to distinguish noise from discrete chaotic maps, such as the logistic map or bifurcation diagram.

In this context , \pecc\ uses a discretized \tser\ through a sampling scheme (described in Section~\ref{subsec:ord-patts}) and calculates the \hentropy\ and \ccomplexity\ values in order to determine what type of behavior (regular, stochastic, complex) is exhibited.
This is done by extracting and counting the occurrence frequency of the sampled data, which are called ``ordinal patterns." Ordinal patterns are groups of points that are ordered from smallest to largest relative amplitude.  The resulting order of indices is that ordinal pattern. For example, if a series of points had values [8, 3, -2, 5] the resulting pattern would be ``3241" since the third value of the array is the smallest, the second value is the second smallest, etc. Section~\ref{subsec:ord-patts} gives a more in-depth discussion of how these ordinal patterns are extracted and determined.

\pecc\ operates on the principle that ordinal patterns may be found within any \tser\ that has $N$ discrete, sequential measurements, calculations, or simulated quantities taken at fixed separation. Since the ordinal patterns are determined purely by comparing relative amplitudes, \pecc\ is agnostic to the physics and other parameters of the system (which often factor into other chaos/noise differentiation methods).

\subsection{Determination of Ordinal Patterns}\label{subsec:ord-patts}
In their pioneering work, \cite{Bandt2002} developed the \hentropy\ ($H$) measure as a means to identify chaotic behavior. Their approach relied on what they called ``ordinal patterns." An 
ordinal pattern is defined as the order in which a subset of $n$ sequential, discrete measurements from a given \tser\ appears such that their values increase from lowest relative amplitude to highest relative amplitude. In cases where there exist two equal values, the original order of points in the \tser\ is preserved. The values themselves are irrelevant since the magnitude of change between steps plays no role in this analysis.

\begin{figure}
\centering
	\includegraphics[width=\columnwidth]{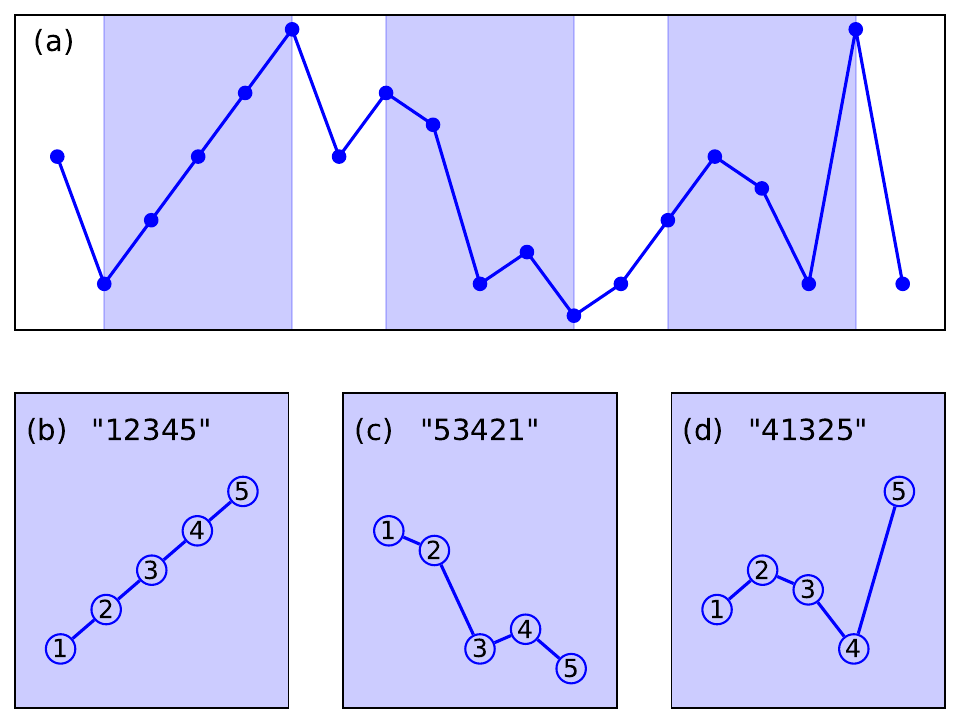}
    \caption{An arbitrary \tser\ of $N=19$ discrete points used to extract ordinal patterns using \edim\ $n=5$.  Shaded regions in panel (a) indicate three examples of patterns using \edel\ $\edelsymbol=1$. Shaded points in panels (b)-(d) show the extracted points with their index numbers labeled on their markers. The corresponding ordinal pattern is listed at the top of each panel.}
    \label{fig:timeseries}
\end{figure}

\begin{figure}
\centering
	\includegraphics[width=\columnwidth]{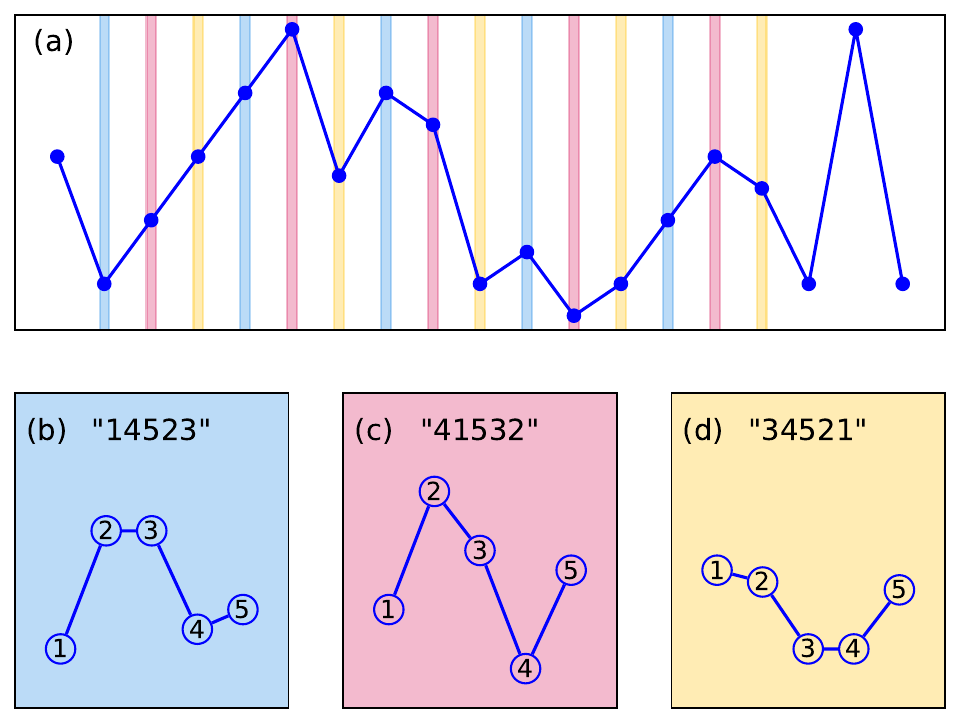}
    \caption{The same arbitrary \tser\ of $N=19$ discrete points from Figure~\ref{fig:timeseries}, but with color-coded shading corresponding to embedding dimension $\edelsymbol=3$ (i.e.,~skipping every two \tstep s). Each highlighted pattern is projected in panels in (b)-(d) to illustrate the ordinal pattern extracted from the $\edelsymbol=3$ sampling. The ordinal patterns extracted using $\edelsymbol=3$ are different than those using $\edelsymbol=1$ in Figure~\ref{fig:timeseries}. }
    \label{fig:timeseries2}
\end{figure}

Figures~\ref{fig:timeseries} and~\ref{fig:timeseries2} illustrate this definition. In Figure~\ref{fig:timeseries}(a), a set of $N=19$ points is shown representing an arbitrary \tser\ along the horizontal axis. The three shaded regions highlight sets of $n=5$ \tstep s, which in this case are both sequential and contiguous. These sequences are again shown in Figures~\ref{fig:timeseries}(b), (c), and~(d), with the ordinal pattern written at the bottom of each panel. The ordinal pattern for these sets of five points is found by first determining which ordinal position has the lowest value, then which ordinal position has the next lowest value, and so on through each of the five points. This pattern can be represented by the numerical sequence shown at the bottom of each of the lower panels in Figure~\ref{fig:timeseries}.

Ordinal patterns are extracted through a method that uses two parameters: the \edim\ $n$ and the \edel\ $\edelsymbol$. The \edim\ $n$ is the number of sequential points extracted to construct the ordinal patterns. Any \tser\ can be decomposed into consecutive, overlapping sets of $n$ \tstep s, where the number of possible permutation orders is $n!$.
The \edel\ $\ell$ is the integer number of \tstep s from one sampled point to the next and probes the corresponding physical \tscal\ in \pecc\ \citep[e.g.,][]{Zunino12,Gekelman14,Weck2015}. 
The \tscal\ for an extracted ordinal pattern associated with a given \edim\ and \edel\ (known as the ``\pattscal") is given by 
    \begin{equation}\label{eq:tpat}
        \tpat=\edelsymbol\, \delta t (n-1)\;,
    \end{equation}
where $\delta t$ is the \tstep\ resolution and the pattern length spans $n-1$ \edel s $\ell$.

The \edel\ for consecutive points is given by $\edelsymbol=1$. It is not necessary, and often not optimal, for the $n$ extracted \tstep s used to construct an ordinal pattern be contiguous (i.e., $\edelsymbol=1$). Figure~\ref{fig:timeseries2} illustrates how ordinal patterns using \edel\ $\edelsymbol=3$ are constructed using the same pattern from Figure~\ref{fig:timeseries}. Here, every third point is grouped as shown by a given color. These three sets of five points are shown in Figure~\ref{fig:timeseries2}(b)-(d) and the numerical representations of their ordinal patterns are also indicated, as in Figure~\ref{fig:timeseries}.

Note that \cite{Bandt2002} called the \edim\ the ``embedding dimension" and the \edel\ the ``embedding delay." These refer to the same parameters, but this paper adopts different language for a more intuitive framing.

In order for \pecc\ to produce meaningful results, the \edim\ $n$ must be large enough that the set of possible ordinal patterns can be robustly used to describe the time-series \citep[$n{>}2$,][]{Bandt2002}, but not so large that the number of patterns becomes intractable.  To the second point, \pecc\ requires the condition $n! \ll N$ be met, where $N$ is the total number of sequential points in the time-series \citep{Rosso07}. \cite{Bandt2002} noted that a practical choice lies in the range $3\leq n \leq 7$. They did not attempt a thorough proof; rather, they showed that chaotic behavior is well identified (even for noisy data) using values of $n$ within these bounds, with a slightly clearer signal near $n=6$. A \edim\ of $n=5$ is adopted throughout this study, following the practice of several studies that effectively use \hentropy~\citep[e.g.,][]{Cao2004,Rosso07,Weck2015}. A more thorough theoretical treatment is beyond the scope of this work but would yield useful justification for one's choice of $n$ in future studies.

\subsection{Pattern Probability and Pattern Probability Distributions}
\label{subsec:probs-dists}
After extracting the $\edelsymbol (n-1)$ ordinal patterns from the \tser, the pattern probability distribution $P$ (also called the ``pattern distribution" by \citet{Weck2015}) can be produced for the $n!$ possible patterns. The probability $p(\pi_i)$ for each pattern $\pi_i$ in $P$ is found by normalizing the occurrence frequency of that pattern so that
    \begin{equation}
        \sum_i^{n!} p(\pi_i) = 1\;,
    \end{equation}
where the subscript $i$ denotes one of the $n!$ possible patterns. The nature of a \tser\ as regular, stochastic, or complex can be discerned by calculating two statistical measures, the \hentropy\ $H$ and \ccomplexity\ $C$ of the resulting pattern probability distribution $P$, for a given \edim\ $n$ and \edel\ $\edelsymbol$. Section~\ref{subsec:H-and-C} introduces and describes these measures in depth.

\begin{figure*}
\centering
	\includegraphics[trim={1.2in 1.2in 0.8in 1.5in},clip, width=\textwidth]{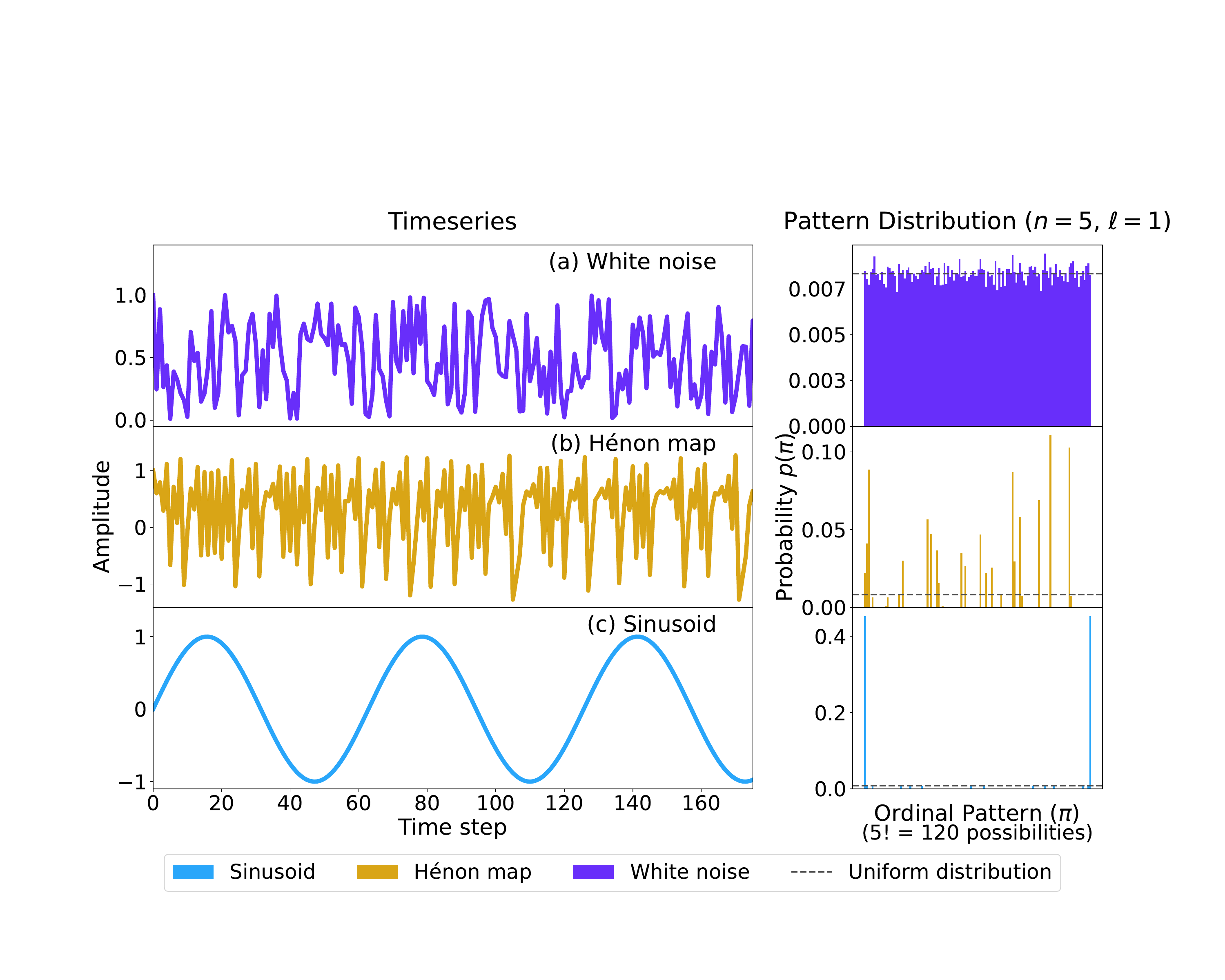}
    \caption{\textit{Left:} Sample section of \tser\ from $t=0$ to $t=175$ out of $t=0$ to $t=5 \times 10^4$ for (a) white noise (stochastic), (b) \henon\ Map (chaotic), and (c) sinusoidal (periodic), where the inset is a zoom-in of a short \tscal\ segment of the series.  
    \textit{Right:} Ordinal pattern probability $p(\pi)$ for possible ordinal patterns $\pi$ for each \tser\ given an \edim\ $n=5$ and \edel\ $\edelsymbol = 1$. The stochastic \tser\ has a uniform distribution of patterns, the periodic \tser\ has a small number of preferred patterns, and the chaotic \tser\ has a variety of preferred, underpreferred, and forbidden patterns. 
    }
    \label{fig:distributions}
\end{figure*}

It is useful to consider the following two extreme cases: (1) a distribution where every pattern is equally represented (e.g., white noise), as in Figure~\ref{fig:distributions}(a), and (2) a periodic \tser\ dominated by very regular patterns (e.g., a sine wave), as visualized by a histogram of the occurrence frequency in Figure~\ref{fig:distributions}(c). Most distributions have a more complex set of occurrence frequencies, as exemplified by the distribution in Figure~\ref{fig:distributions}(b). Such distributions reveal preferred (high occurrence frequency) and/or forbidden patterns (low or zero occurrence frequency).

Time-series data analyzed using \pecc\ must adequately populate the $n!$ patterns in order to ensure the value for each occurrence probability $p(\pi_i)$ is statistically significant. In practice, either the \tser\ must be sufficiently long or multiple \tser\ can be combined. The first case is appropriate for long \dset s where the characteristic behavior of the \tser\ is not time-dependent, as in the case study in \S\ref{subsubsec:exs-dpen}.  The latter is appropriate for shorter characteristic \tscals\ and requires an ensemble of \tser. In any case, \pecc\ is only able to probe the characteristic behavior of a \tser\ at \tscals\ corresponding to an appropriately sampled time domain. Section~\ref{subsec:sampling-limits} discusses a method for determining the minimum \tser\ duration as well as a range of appropriate sampling intervals.

\subsection{Permutation Entropy and Statistical Complexity}\label{subsec:H-and-C}
The core of the \pecc\ method is the calculation of the \hentropy\ and \ccomplexity, which are used in combination to evaluate whether a given \tser\ is regular, complex, or stochastic. Table~\ref{table:stats-terms} provides a glossary of the terms used in this section, as well as equation references.

\subsubsection{Permutation Entropy}\label{subsubsec:H}
A common metric for a pattern probability distribution $P$ is the Shannon entropy (or information entropy) \citep{shannon:1948}, expressed as
    \begin{equation}\label{eq:S}
        S[P] = -\sum_i^{n!} p(\pi_i) \log p(\pi_i) \;.
    \end{equation} 
The value of $S$ normalized to its maximum possible value, i.e.,
    \begin{equation}\label{eq:H}
        H[P] = \dfrac{S[P]}{\log n!},
    \end{equation}
is the \hentropy\ \citep{Bandt2002,Rosso07}. Using this metric, a \tser\ that is dominated by a single pattern (e.g., a linear ramp), would have $H= 0$, while an equally probable distribution of patterns (e.g., white noise, as in Figure~\ref{fig:distributions}(a)), would have $H=1$. An intermediate pattern, as in Figure~\ref{fig:distributions}(b), would have an intermediate value for $H$.

Periodic time-series, such as sine waves or triangle waves, have limited numbers of possible permutations, as in Figure~\ref{fig:distributions}(c). There are upward and downward ordered ramp patterns and some additional patterns from permutations that include local maxima or minima in the periodic function.

The number of possible patterns does not change with embedding delay since the same patterns are possible no matter the sampling resolution. However, the probability for a given pattern does depend on the sampling resolution. For example, the probability for ramp patterns increases as the \edel\ decreases.

The lower limit for the value of $H$ for a periodic function is the limit when only two ramping patterns (upward and downward) are measured.  That is,
    \begin{equation}\label{eq:Hper_min}
        \HMinPer(n)=\frac{\log 2}{\log(n!)}.
    \end{equation}
The limiting minimum possible value for the \hentropy\ of a periodic function when $n=5$ is thus $\HMinPer(n{=}5)=0.14478$.

The limiting maximum possible value of $H$ for a periodic function is the case when the probability of each possible pattern is equal. The number of possible patterns for any periodic function composed of two ramping patterns is conjectured to be
    \begin{equation}\label{eq:periodicpatterns}
        N_{\rm periodic}(n)=2[2(n-2)+1].
    \end{equation}
This can be understood in the case of a triangle function. There is one upward ramping pattern (last term in brackets), and there are $(n-2)$ non-ramp patterns around the crest where any evenly spaced sampling will either have points sampled on the right side  staggered with higher than values than those on the left or vice versa (second term in brackets). There is also a symmetry for downward ramping points and points around the trough. The same ordinal patterns exist for single-frequency periodic functions, like sine and cosine, which are indistinguishable from a triangle function with the same maximum/minimum frequency. This hypothesis has been tested for the range of \edim s\ $3<n<8$. It follows that
    \begin{equation}\label{eq:Hper_max}
        \HMaxPer(n)=\frac{\log N_{\rm periodic}(n)}{\log(n!)}.
    \end{equation}
Since $N_{\rm periodic}(n{=}5)=14$, a periodic function sampled with a \edim\ of $n{=}5$ is expected to have $H{\leq} \HMaxPer{=}0.55124$.

\subsubsection{Disequilibrium and Complexity}\label{subsubsec:D-and-C}
The pattern probability distribution $P$ can also be characterized by how poorly it is described by the pattern probability distribution for the uniform case, $P_e$, where each possible pattern has equal probability $p_e=1/n!$. The divergence of ensemble $P$ from $P_e$ is called the ``disequilibrium" and is defined as
    \begin{equation}\label{eq:disequilibrium}
        d[P,P_e] = S \left[ \frac{P+P_e}{2} \right] - \frac{1}{2}S[P]-\frac{1}{2}S[P_e] \;,
    \end{equation}
where $S[P+P_e]$ is the Shannon entropy for the sum of pattern probability distributions $P$ and $P_e$. The value of the disequilibrium $d[P,P_e]$ is normalized by its maximum possible value ($d/d_{\rm max}$) is given by \citep{Lamberti04}
    \begin{equation}\label{eq:D}
        D[P,P_e] = \dfrac{2d[P,P_e]}{2 \log(2n!)-\log(n!)-\frac{n!+1}{n!} \log(n!+1)} \;,
    \end{equation}
and scales in the opposite direction to the \hentropy\ (e.g.,~$D\rightarrow1$ as $H\rightarrow 0$). \Ccomplexity, also known as the Jensen-Shannon complexity, is given by the product~\citep{Lamberti04,Rosso07} 
    \begin{equation}\label{eq:C}
        C[P,P_e] = D[P,P_e]\,H[P].
    \end{equation}
Low values for $C$ indicate a system with a distribution of ordinal patterns that is either far from the uniform distribution, as $H$ approaches zero, or very near the uniform distribution, as $D$ approaches zero. Maximum complexity occurs in an intermediate range when both $H$ and $D$ are nonzero.

The \ccomplexity\ ($C$) measure has been compared to several established methods (e.g., \Lyp\ analysis) and emerging ones \citep[e.g.,~the LMC measure of][]{LMC95} for identifying chaotic or complex behavior and is shown to be robust for a wide range of scenarios \citep{Lamberti04,Rosso07}, including logistic maps, the skew tent map, the \henon\ map, the Lorenz map of Rossler's oscillator, and Schuster maps \citep{Schuster88}.
Further, $C$ is an intensive measure that can be used to provide insight into the dynamics of a system \citep{Lamberti04}, such as relevant \tscals\ (explored in Sections~\ref{subsec:HC-plane} \& \ref{sec:examples}). It is also reliably able to quantify the degree of chaos in systems that also have some degree of periodicity \citep{Lamberti04} or \stocity\ \citep{Rosso07,Zunino12}.

\begin{deluxetable*}{Lllc}
    \label{table:stats-terms}
	\tablecaption{Glossary of Statistical Terms}
	\tablehead{\colhead{Symbol} & \colhead{Name} & \colhead{Definition} &
    \colhead{Eq. no} }
    \startdata
        P &
            \makecell[l]{Pattern probability\\distribution} &
            \makecell[l]{All possible $n!$ ordinal pattern permutations} &
            \S\ref{subsec:probs-dists} \\
        P_e &
            \makecell[l]{Equilibrium pattern\\ probability distribution} &
            \makecell[l]{Uniform distribution of all possible\\ $n!$ ordinal pattern permutations} &
            \S\ref{subsubsec:D-and-C} \\
        \pi_i &
            $i$-th ordinal pattern &
            \makecell[l]{A possible pattern permutation of the pattern\\ probability distribution $P$} &
            \S\ref{subsec:probs-dists} \\
        S &
            Shannon entropy &
            \makecell[l]{Information entropy} &
            \S\ref{subsubsec:H}, Eq.~\ref{eq:S} \\
        H &
            Permutation entropy &
            \makecell[l]{Normalized Shannon entropy, measure used\\ in \hcplane\ analysis} &
            \S\ref{subsubsec:H}, Eq.~\ref{eq:H} \\
        \HMinPer(n) &
            \makecell[l]{Minimum possible\\ \hentropy\\ for a periodic function} &
            \makecell[l]{Smallest value of $H$ for a periodic function\\ (e.g., sine wave); dependent on \edim} &
            \S\ref{subsubsec:H}, Eq.~\ref{eq:Hper_min} \\
        \HMaxPer(n) &
            \makecell[l]{Maximum possible\\ \hentropy\\ for a periodic function} &
            \makecell[l]{Largest value of $H$ for a periodic function\\ (e.g., sine wave); dependent on \edim} &
            \S\ref{subsubsec:H}, Eq.~\ref{eq:Hper_max} \\
        \Hofl &
            \Hcurve &
            \makecell[l]{\Hentropy\ as a function of the\\ \edel} &
            \S\ref{subsec:sampling-limits} \\
        d &
            Disequilibrium &
            \makecell[l]{Measure of how far pattern probability \\ distribution $P$ is from a uniform\\ distribution of patterns} &
            \S\ref{subsubsec:D-and-C}, Eq.~\ref{eq:disequilibrium} \\
        D &
            Normalized disequilibrium &
            \makecell[l]{Normalized measure of disequilibrium used\\ in calculation of $C$} &
            \S\ref{subsubsec:D-and-C}, Eq.~\ref{eq:D} \\
        C &
            \makecell[l]{Jensen-Shannon\\ \ccomplexity} &
            \makecell[l]{Measure used in \hcplane\ analysis} &
            \S\ref{subsubsec:D-and-C}, Eq.~\ref{eq:C} \\
        \Cofl &
            \Ccurve &
            \makecell[l]{\Ccomplexity\ as a function of\\ the \edel} &
            \S\ref{subsec:sampling-limits} \\
	\enddata
\end{deluxetable*}

\subsection{The $HC$-Plane}\label{subsec:HC-plane}
Any \tser\ can be qualitatively sorted into its degree of regular, stochastic, and/or complex behavior by combining the metrics for the \hentropy\ $H$ (normalized Shannon entropy) 
and \ccomplexity\ $C$ (normalized Jensen-Shannon complexity) for a given \edel\ $\edelsymbol$ and \edim\ $n$ \citep{Rosso07}. Specifically, $H$ provides a quantitative scale for how stochastic or noisy the \tser\ is, while $C$ measures the degree of complexity or chaos by how many statistically preferred and/or forbidden patterns there are.

\begin{figure}
    \centering
    \includegraphics[width=\columnwidth]{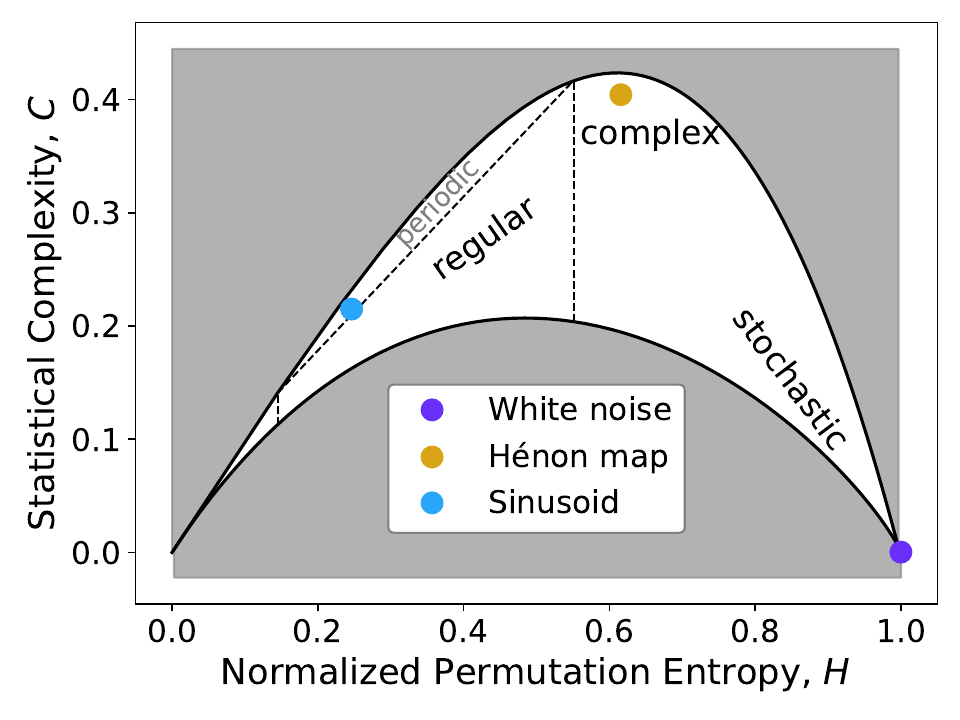}
    \caption{The \hcplane\ is an effective visualization of \hentropy\ and \ccomplexity, where the value for the \hentropy\, $H$, for a \tser\ assuming given \edel\ is plotted on the $x$-axis and the value for \ccomplexity\, $C$, is plotted on the $y$-axis. The upper and lower limits for $C$ are indicated by the solid (black) crescent-shaped curves, here specifically for $n=5$. The relative scale of these boundaries depends on the embedding dimension, though it varies only slightly from $n=3$ through $n=6$. The regions associated with \stocity\ fall in the lower right, while those associated with \cmplx\ behavior fall in the upper central part of the plane. The diagonal dashed line indicates the boundary for the region for a purely periodic function, with the regular orbits outside of it. The $\hcval$ coordinates are shown for the \tser\ in Figure~\ref{fig:distributions} with \edim\ $n=5$ and \edel\ $\edelsymbol =1$. Vertical dashed lines represent the minimum and maximum \hentropy\ values for a purely periodic function (i.e., $\HMinPer$ and $\HMaxPer$).
    }
    \label{fig:HCplane_diagram}
\end{figure}

These regimes can be visualized as locations on a coordinate plane where the \hentropy\ ($H$) is on the $x$-axis and the \ccomplexity\ ($C$) is on the $y$-axis. Figure~\ref{fig:HCplane_diagram} illustrates the \hcplane\ with maximum and minimum limiting values for $C(H)$ indicated with solid curves, where the gray regions outside these curves are forbidden. The bounding curves are computed following a technique from ~\cite{Calbet2001} using a Lagrange multiplier technique for each fixed \hentropy\ from Equation~\ref{eq:C}. Regular \tser\ generate coordinates that occupy the left-hand region of the \hcplane\ while noisy \tser\ occupy the lower right. Complex or chaotic \tser\ occupy the upper middle region {\citep{Rosso07,Zunino12,Gekelman14,Weck2015}}. Purely periodic functions (see discussion in \S\ref{subsubsec:exs-sine}) and circular orbits (see \S\ref{subsubsec:exs-kep}) fall on or within the region bounded by the diagonal dashed boundary line.

There are three example \tser\ plotted on the \hcplane\ in Figure~\ref{fig:HCplane_diagram}. These \tser\ are shown in Figure~\ref{fig:distributions}, where sampling parameters of \edel\ $\edelsymbol = 1$ and \edim\ $n=5$ were used to produce each $\hcval$ coordinate in Fig.~\ref{fig:HCplane_diagram}. The \tser\ generated from a uniform random number generator shown in Figure~\ref{fig:distributions}(a) (purple) is labeled as white noise and occupies the most extreme lower right position in the \hcplane. 
The \henon\ map is a system described by \citet{Henon76},
\begin{equation}\label{eq:Henon}
  (x_m,y_m)=\begin{cases}
    x_{m+1} = 1-ax_m^2 + y_m\\
    y_{m+1} = bx_m
  \end{cases}\;,
\end{equation}
which produces the chaotic \tser\ shown in Figure~\ref{fig:distributions}(b) (yellow) using the selected parameters, $a=1.4$ and $b=0.3$. The $\hcval$ coordinate from this \tser\ lies near the very top of the complexity region. The sine wave shown in Figure~\ref{fig:distributions} falls on the boundary line of the purely periodic region on the left side of the \hcplane.

\section{Usage and Interpretation}\label{sec:usage-interpretation}
\subsection{Setting up \pecc}\label{subsec:pecc-setup}
To use \pecc, the code can be installed from the Python Package Index (PyPI) via the command \texttt{pip install peccary} or downloaded from the GitHub repository \citep{peccaryZenodo}.\footnote{\url{https://github.com/soleyhyman/peccary}} Documentation and tutorials for running the code can be found on the \pecc\ website.\footnote{\url{https://peccary.readthedocs.io}} At its most basic, all that is needed is a \tser\ and a chosen sampling interval $\edelsymbol$. By default, the \edim\ is set to $n = 5$ (see Section~\ref{subsec:ord-patts}). 

Typically, the \tser\ measures used are determined by the system and the symmetry in question. For example, when investigating orbital behavior in a barred disk, the appropriate choice may be the Cartesian coordinate along the length of the bar in the rotating frame to discern the behavior of those orbits.

\subsection{Idealized Sampling Scheme and Limitations}\label{subsec:sampling-limits}
Due to the flexibility of the \pecc\ method, it is possible to probe the orbital behavior on a variety of different \tscals. This can be done by calculating $H$ and $C$ for a range of different \edel s $\edelsymbol$ and producing \HandCcurves, or $\Hofl$ and $\Cofl$. Alternatively, a single \pattscal\ or \edel\ can be chosen to probe the \tscal\ of maximum \ccomplexity\ or any generic \tscal. However, if the chosen \edel, $\edelsymbol$, is poorly matched with the natural \tscal\ of the system, or the overall duration of the \tser, $\tdur$, is insufficient, the interpreted results may be inaccurate. For example, if a continuous chaotic \tser\ is sampled at small enough intervals, ramping behavior will dominate. Similarly, if the same \tser\ is sampled with too large a \edel, it will appear stochastic.

Any given \tser\ has three primary \tscals\ of interest. These are the overall duration of the \tser\ $\tdur$, the natural \tscal\ of the system $\tnat$, and the \pattscal\ $\tpat$ of the ordinal pattern sampling scheme. One can find ratios to relate these \tscals\ to one another. Below is a description of the method adopted by this study to guide in the selection of appropriate parameters for a given \tser\ that is based on these ratios. Table~\ref{table:samps-tscals} lists the relevant \tscals\ and sampling parameters, their definitions, and references to their descriptions in this text.

The number of natural \tscals\ (e.g.,~orbital periods) in a given \tser\ is represented by the ratio $\tdurtnat$. The time resolution required to capture the nature of the \tser\ can be represented by the ratio $\tpat/\tnat$. 

Systems with well-known periodic or chaotic behavior are used to determine the minimum necessary constraints for $\tdurtnat$ and $\tpat/\tnat$.

\subsubsection{ Minimum \Tser\ Duration, $\tdurtnat$}\label{subsubsec:minTdurTnat}
Several \tser\ with a range of durations, $\tdur$, were created for a sine wave with given fixed period, where $t_{\rm period}=\tnat$.  These were used to determine the minimum duration necessary to diagnose a given system, $\tdurtnat$. Ratios ranged between $\tdurtnat=0.5 - 10$. 
For each of these \tser, values for $\hcval$ were calculated for a range of selected $\tpat$ such that $\tpattnat=0.1-0.7$. The resulting $\hcval$ values were then plotted on the \hcplane\ and as $H(\tpattnat)$ curves.

Purely periodic/closed functions such as sine waves have a characteristic behavior in their \Hcurve s in that they increase from the lower limit of $\HMinPer$ (at small \edel s) until they reach the upper limit of $\HMaxPer$ and then oscillate between $\HMaxPer$ and lower values of $H$. On the \hcplane, this corresponds to the $\hcval$ points falling exactly on the periodic boundary line or zigzagging between that boundary line shown in Figure~\ref{fig:HCplane_diagram} and the region to the left of it. 

Within the range of \tser\ durations sampled, $\hcval$ values diverged to the right of the periodic boundary and did not reach the $\HMaxPer$ upper limit when $\tdur$ fell below critical thresholds. For the sine wave, $H(\tpattnat)$ stopped reaching $\HMaxPer$  at $\tdurtnat{\sim}1.5$, while the \hcplane\ behavior significantly deviated from the aforementioned characteristic behavior at $\tdurtnat{\sim} 1$. The first two rows of Figure~\ref{fig:samplingConstraints} illustrate this behavior.

In cases where the duration of the orbital behavior in question is shorter than this limit, one might consider stacking \tser\ for multiple orbits. Initial explorations indicate that stacking multiple, shorter-duration \tser\ can return reliable results. This will be further explored in a later paper in this series.

\subsubsection{Timescale Resolution, $\tpattnat$}\label{subsubsec:maxTpatTnat}
To identify the largest value of $\tpattnat$ that should be used with \pecc, the same $H(\tpattnat)$ plots created for identifying the minimum $\tdurtnat$ (Section~\ref{subsubsec:minTdurTnat}) were used. To establish a conservative upper limit, the maximum $\tpattnat$ was found by locating the lowest value of $\tpattnat$ at which $H(\tpattnat)$ for a sine wave fell significantly below the $\HMinPer$ line, regardless of the use of an appropriate $\tdurtnat$ ratio. For the sine wave, this occurred at $\tpattnat {\sim} 0.5$ when $\tdurtnat{\sim} 0.6$. The third row of Figure~\ref{fig:samplingConstraints} shows this graphically.

For the lower limit of $\tpattnat$, the $x$-coordinate \tser\ from the chaotic Lorenz strange attractor simulation were used. Similar to the processes used to constrain $\tdurtnat$ and to establish an upper limit for $\tpattnat$, \Hcurve s and \hcplane\ plots were generated for a range of $\tpattnat$ values, ranging from 0.1 to 0.7 with the $\HMinPer$ and $\HMaxPer$ lines overplotted. The minimum $\tpattnat$ was set to be the value at which the \Hcurve\ crossed the $\HMaxPer$ line (i.e., transitioning from appearing regular to appearing complex). For the $x$-coordinate \tser\ for the Lorenz strange attractor with $\tdurtnat = 1.5$, this occurred at $\tpattnat {\sim} 0.25$. For a more conservative constraint, this was rounded up to $0.3$. The fourth row of Figure~\ref{fig:samplingConstraints} illustrates this.

\subsubsection{Recommended Sampling Scheme Constraints}\label{subsubsec:samplingConstraints}

The sampling scheme tests performed in Sections~\ref{subsubsec:minTdurTnat} and \ref{subsubsec:maxTpatTnat} used two systems with known behavior, i.e., a periodic (sinusoid) function and a continuous chaotic system (Lorenz strange attractor). To obtain reliable $\hcval$ values, the \tser\ duration must be at least of order of the natural \tscal\ (i.e., $\tdur/\tnat > 1$) and preferably $\tdur/\tnat {\gtrsim} 1.5$, and the time resolution should fall in an approximate range of $0.3 \lesssim \tpat/\tnat \lesssim 0.5$.  In practice, this ratio can be used to select an appropriate value for \edel\ $\edelsymbol$.

\begin{figure*}
    \centering
    \includegraphics[width=0.8\textwidth]{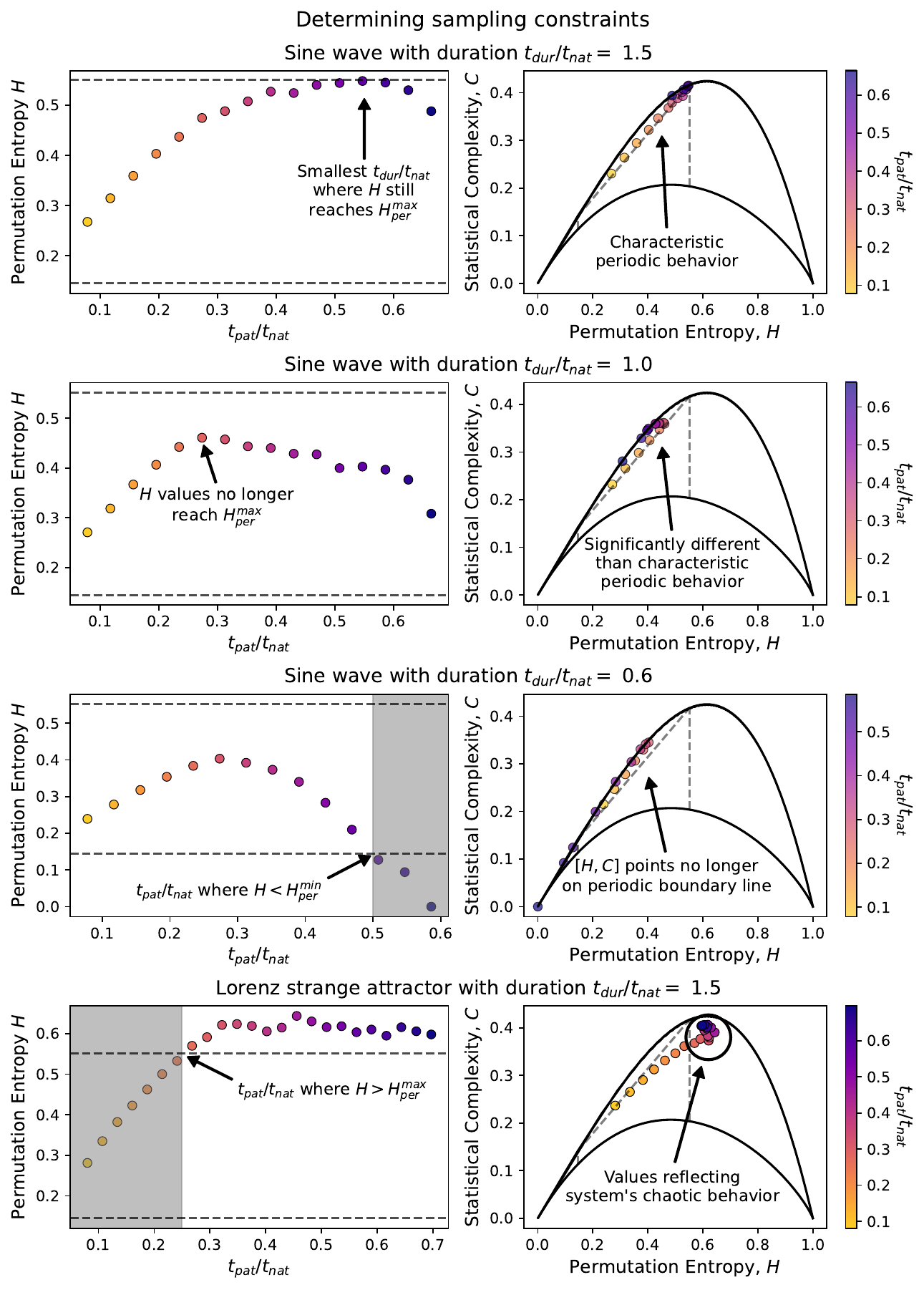}
    \caption{Illustration of method diagnostic for identifying sampling scheme constraints. \textit{Top row:} $H(\tpattnat)$ and \hcplane\ plots for a sine wave \tser\ with a duration of $\tdurtnat = 1.5$, with $\tnat$ being the period of the sinusoid. The \hcplane\ on the right demonstrates the characteristic behavior of a periodic function, and the \Hcurve\ plot on the left shows that $H(\tpattnat)$, or $\Hofl$ does not reach the $\HMaxPer$ upper limit. \textit{Second row:} $H(\tpattnat)$ and \hcplane\ plots for a sine wave \tser\ with $\tdurtnat = 1$, i.e., $\tdur = \tnat$. Both \Hcurve\ and \hcplane\ plots show that behavior deviates significantly from the characteristic behavior. \textit{Third row:} $H(\tpattnat)$ and \hcplane\ plots for a sine wave \tser\ with $\tdurtnat = 0.6$. The \Hcurve\ shows the value for $\tpattnat$ at which $H < \HMinPer$. Points on \hcplane\ do not fall on the periodic boundary line. \textit{Fourth row:} $H(\tpattnat)$ and \hcplane\ plots for a chaotic Lorenz strange attractor \tser\ with $\tdurtnat = 1.5$. The \Hcurve\ shows the location at which $H(\tpattnat) > \HMaxPer$, indicating the region where the chaotic \tser\ is reliably classified as complex. A circle on the \hcplane\ marks the accurately classified points.}
    \label{fig:samplingConstraints}
\end{figure*}

Note that all of these limits are derived using a \edim\ of $n=5$ and a similar process will need to be followed in order to find the appropriate constraints when using other values for $n$. 
Figure~\ref{fig:samplingConstraints} shows example diagnostic plots used to obtain the constraints reported in this paper.

\begin{deluxetable*}{lLllc}
    \label{table:samps-tscals}
	\tablecaption{Glossary of Sampling Terms and \Tscals}
	\tablehead{\colhead{Parameter Type} & \colhead{Symbol} & \colhead{Name} & \colhead{Definition} &
    \colhead{Eq. no}}
    \startdata
        \multirow{2}{*}{Sampling} & n &
            \Edim &
            \makecell[l]{Number of discrete points sampled for a pattern} & 
            \S\ref{subsec:ord-patts} \\
         & \edelsymbol &
            \Edel &
            \makecell[l]{Number of points spanned by each sample within\\ a pattern} &
            \S\ref{subsec:ord-patts} \\
        \hline 
        \multirow{4}{*}{Timescales} & \delta t &
    	    \Tstep &
    	    \makecell[l]{Time element associated with a single step\\ in the \tser} &
    	    \S\ref{subsec:ord-patts} \\
         & \tpat & 
    	    \Pattscal &
    	    \Tscal\ for an ordinal pattern &
    	    \S\ref{subsec:ord-patts}, Eq.~\ref{eq:tpat} \\
         & \tdur &
            \Tser\ duration &
            Total duration of a \tser &
            \S\ref{subsec:sampling-limits} \\
         & \tnat &
    	    Natural \tscal &
    	    \makecell[l]{Natural or approximate period of oscillation\\ for the system} &
    	    \S\ref{subsec:sampling-limits} \\
	\enddata
\end{deluxetable*}

\subsection{Interpreting \pecc\ Values}\label{subsec:interpretation}
There are two methods by which one can interpret the values of \hentropy\ and \ccomplexity\ produced by \pecc: using \HandCcurves\ or by calculating a single value of $\hcval$ at an optimal pattern \tscal. This subsection compares the benefits and drawbacks of each.

\textit{\HandCcurves} --- The most exact way is to plot the \HandCcurves\ for each orbit within the system, which will probe the behaviors of the orbits on different \tscals. Should the system be evolving with time, the \tscal\ of the orbital behavior in question should be used to approximate the duration of the \tser, $\tdur$, when considering whether or not its nature can be discerned using a single orbit with \pecc.

Regular, complex, and stochastic \tser\ will all have different $\Hofl$ and $\Cofl$ shapes. For stochastic \tser, the \hentropy\ and \ccomplexity\ curves are close to constant, with $\Hofl \sim 1$ and $\Cofl \sim 0$. This is due to the fact that generated noise or \stocity\ does not have any characteristic \tscals.

Chaotic systems, on the other hand, have a characteristic shape to their curves that depends on whether they are discrete or continuous in nature. 
Discrete, recursive, chaotic mappings or sequences (such as the \henon\ map), have elements that are labeled with integer indices (e.g., $\{x_0,x_1,\ldots,x_{m-1},x_m\}$). The maximum statistical complexity of such a map will occur at the densest possible sampling interval of $\edelsymbol=1$. By contrast, the maximum \ccomplexity\ for continuous chaotic systems depends on the approximate natural timescale. In terms of \HandCcurves, the value for \hentropy\ increases with increasing \edel, while the value for \ccomplexity\ increases to some maximum value at a particular \pattscal\ and then decreases. Examples of both discrete and chaotic maps are given in Sections~\ref{subsubsec:exs-chaos} and \ref{subsubsec:exs-dpen}.

Compared to stochastic and complex signals, regular \tser\ generally have smaller values for $H(\edelsymbol{=}1)$ that rise with increasing \edel. The \Hcurve s\ of purely periodic \tser\ (such as a sine wave) also exhibit a characteristic pattern of  $H(\edelsymbol) \rightarrow \HMaxPer$ at a ratio of $\tpat/\tnat{\sim}0.6$ and regularly return to that value as the \edel\ continues to increase. This behavior is reflected in the \Ccurve\ as well (since $C$ depends on $H$), which results in a purely periodic function falling on or within the periodic boundary of the \hcplane\ for all \edel\ values. This is further discussed in Section~\ref{subsubsec:exs-sine}.

\textit{Single $\hcval$ value} --- For very large datasets and many particles, generating \HandCcurves\ for each \tser\ can be impractical. In these cases, the next-best method is to choose a \edel\ within the limits for $\tpat/\tnat$, as described in Section~\ref{subsec:sampling-limits}. The locations where the values fall within the \hcplane\ in Figure~\ref{fig:HCplane_diagram} result in the classification of the orbit type.

In ambiguous cases, it may be necessary to incorporate additional methods, such as Fourier analysis in order to break some of the degeneracy/uncertainty. This will be the subject of a future paper in this series. See Section~\ref{sec:future} for further discussion.

\section{Examples}\label{sec:examples}
This section provides a variety of examples demonstrating the performance of the \pecc\ method in systems with known outcomes. In Section~\ref{subsec:exs-wellChar}, \pecc\ is applied to several well-characterized, mathematical functions, while Section~\ref{subsec:exs-astro} demonstrates the usage of \pecc\ on tracer particle simulations of three well-known orbital systems.

\subsection{Well-characterized Mathematical Examples}\label{subsec:exs-wellChar}

\subsubsection{Sine Wave}\label{subsubsec:exs-sine}
The sine wave is a classic example of a purely periodic function. This example generates five sine waves of different periods. Each of these \tser\ has a duration of $\tdur=10~\mathrm{s}$, sampled at a resolution of $\delta t = 2^{-8}~\mathrm{s}$, and the duration of each \tser\ is greater than at least five completed cycles ($\tdurtnat > 5$). 

\begin{figure*}
    \centering
    \includegraphics[width=\textwidth]{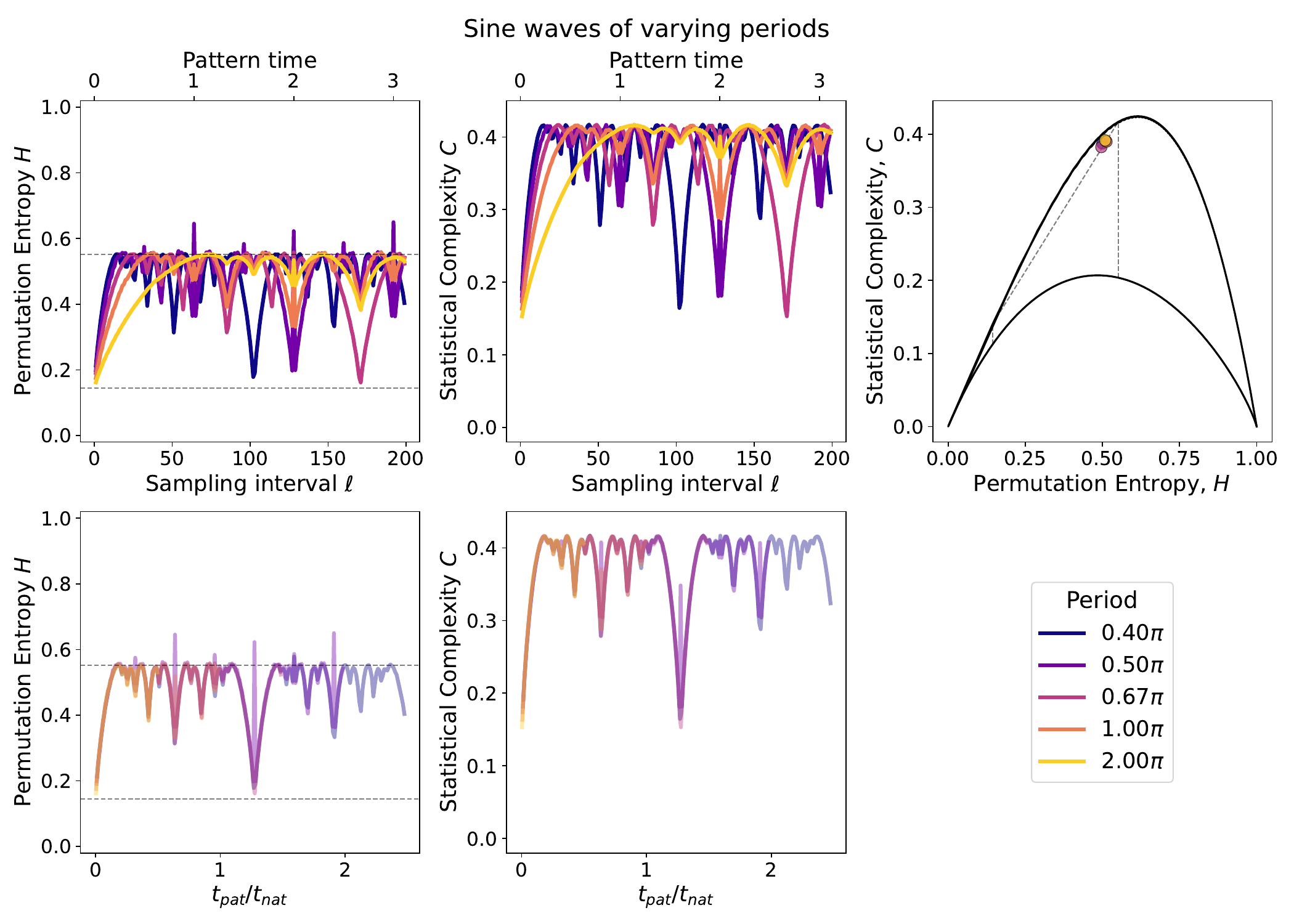}
    \caption{Values of \hentropy\ and \ccomplexity\ values for a range of \edel s for five sine waves of different periods. \textit{Top left}: \hentropy\ values for \edel s ranging from 1 to 200, with the corresponding \pattscal s on the top $x$-axis. Sine waves with shorter periods reach the $\HMaxPer(n=5)$ limit more quickly. Horizontal dashed lines in the two left panels represent the lower and upper limits of $\HMinPer(n=5)$ and $\HMaxPer(n=5)$, respectively. \textit{Bottom left}: The same \hentropy\ values except that the $x$-axis is the \pattscal\ scaled by the period of each sine wave (i.e., $\tpattnat$). All five of the \Hcurve s overlap exactly, with the exception of the numerical spikes. \textit{Top center}: \ccomplexity\ values as a function of \edel/\pattscal. As with the \Hcurve s, the sine waves with shorter periods reach the initial peak much more rapidly than those with longer periods. \textit{Bottom center}: \ccomplexity\ values plotted against the $\tpattnat$ ratio show that the \Ccurve s for the different sine waves have the same functional form. \textit{Top right}: The $H$ and $C$ values for the five sine waves with ideal sampling ($\tpat/\tnat =0.4$, $\tdurtnat \geq 1$) plotted on the \hcplane\ all fall on or within the boundary region for a purely periodic function.}
    \label{fig:sine-ex}
\end{figure*}

Figure~\ref{fig:sine-ex} shows how the values for the \hentropy\ (top left panel) and \ccomplexity\ (top middle panel) depend on the choice of \edel\ ($\edelsymbol$).  The patterns in these plots are clearer when plotting both as a function of the \pattscal\ divided by the period (i.e., natural \tscal), $\tpat/\tnat$ (bottom panels).  These panels illustrate that there is a characteristic shape to the curves of \hentropy\ and \ccomplexity\ that depends on the period of the oscillatory behavior.  The curve for the \hentropy\ has a maximum value set by $H^{\rm max}_{\rm periodic}$ (Equation~\ref{eq:Hper_max}) with the exception of three spikes that are numerical artifacts.  The \hcplane\ (top right panel) shows the distributions for all choices for $\edelsymbol$.

\subsubsection{Noise Varieties}\label{subsubsec:exs-noises}
\begin{figure*}
    \centering
    \includegraphics[width=\textwidth, trim=2cm 0cm 2cm 0cm]{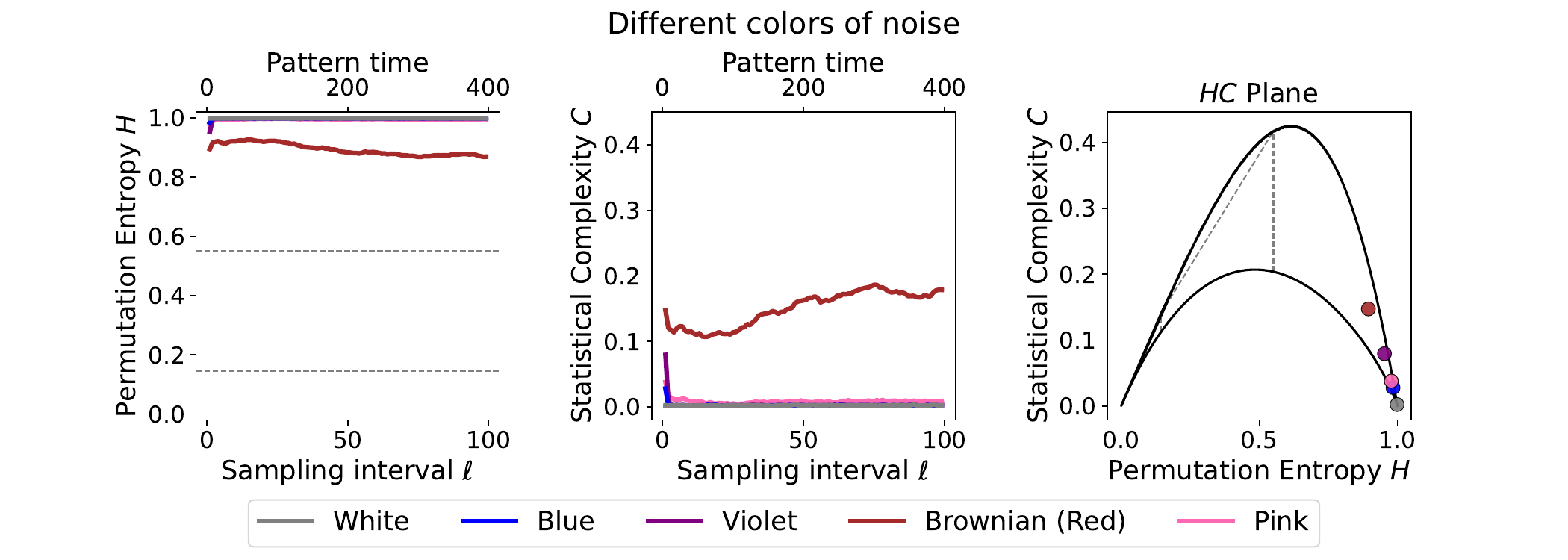}
    \caption{Values of \hentropy\ and \ccomplexity\ values for a range of \edel s for five different colors of noise (pink, red, violet, blue, and white). \textit{Left}: \hentropy\ as a function of \edel/\pattscal. All $H$-values are very high (close to $H=1$), indicating the presence of close to all possible permutations of patterns. Horizontal dashed lines represent the lower and upper limits of $\HMinPer(n=5)$ and $\HMaxPer(n=5)$, respectively. \textit{Center}: \ccomplexity\ as a function of \edel/\pattscal. $C$-values are very low for all colors of noise, indicating that the pattern probability distributions are close to uniform. \textit{Right}: the $\hcval$ values for a \edel\ of $\edelsymbol = 1$ plotted on the \hcplane\ for all noise varieties fall well within the stochastic region. Note that Brownian noise has slightly lower \hentropy\ and slightly higher \ccomplexity\ than some of the other noise colors due to the steep slope of its power spectrum, which results in the suppression of higher frequencies and in less overall scatter in the resulting \tser.}
    \label{fig:noise-ex}
\end{figure*}

\pecc\ is effective for a variety of colors (or power spectra) of noise. 
Figure~\ref{fig:noise-ex} shows the values for the \hentropy\ (left panel) and \ccomplexity\ (middle panel) as a function of the \edel\ ($\edelsymbol$) for five varieties of noise.  These are white noise (power spectral density equal at all frequencies $\nu$), blue noise (power spectral density $\propto \nu$), violet noise (power spectral density $\propto \nu^2$), Brownian noise (also called red noise, power spectral density $\propto \nu^{-2}$), and pink noise (power spectral density $\propto \nu^{-1}$). Using \pecc's \texttt{examples.noiseColors} class, five sample \tser\ of $10^4$ discrete measures were created for the aforementioned noise colors. 

Each noise spectrum has values on the \hcplane\ that are indicative of \stocity. Furthermore, the \hentropy\ and \ccomplexity\ curves (i.e. $\Hofl$ and $\Cofl$) from any type of noise have nearly constant values for all choices for \edel\ $\edelsymbol$, where the value for $C$ is close to 0 and the value for $H$ is close to 1 at all scales. This is due to the fact that the value of the \tser\ at each \tstep\ comes from a random distribution, which means the occurrence frequency of patterns at every \edel\ will always be uniform or very nearly so.

\subsubsection{Chaotic Systems}\label{subsubsec:exs-chaos}
\begin{figure}
    \centering
    \includegraphics[width=\columnwidth, trim=2cm 1cm 0cm 0cm]{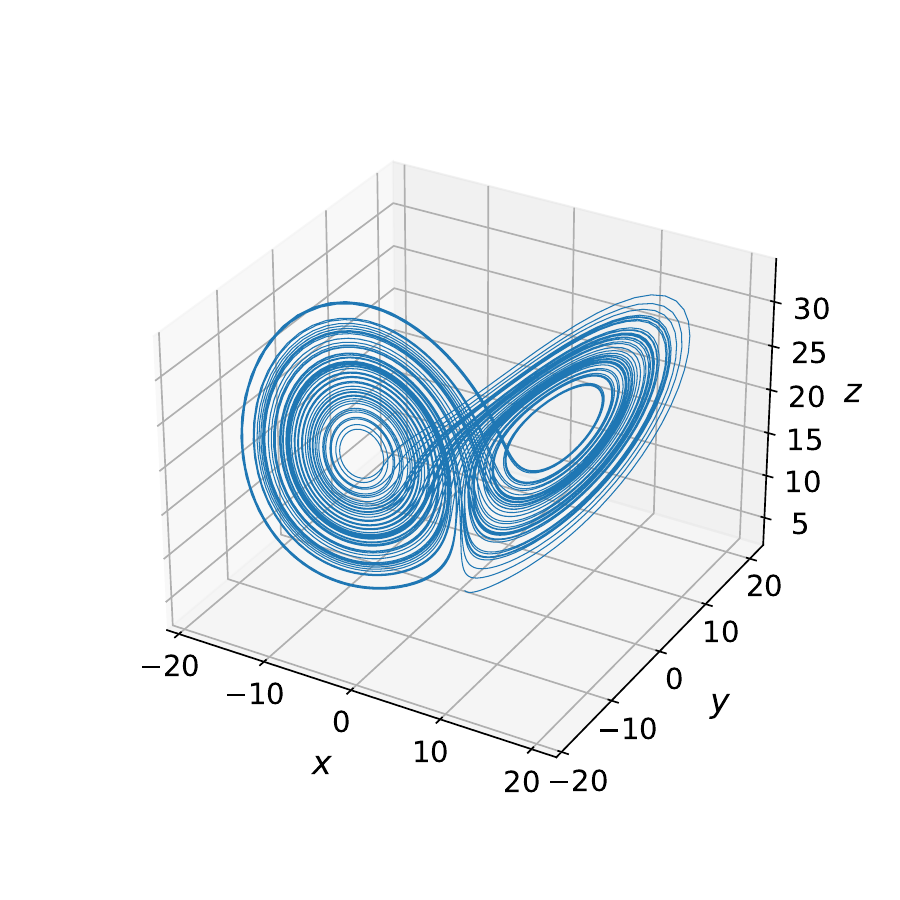}
    \caption{Three-dimensional plot of the Lorenz strange attractor with parameters $\sigma=10$, $\rho = 20$, and $\beta = 2.667$.}
    \label{fig:lorenz-ex}
\end{figure}

Two well-studied examples of chaos are the \henon\ map (Eq.~\ref{eq:Henon}) and the Lorenz strange attractor. The \henon\ map is a discrete chaotic map, while the Lorenz strange attractor is continuous. The Lorenz strange attractor is a system described by \citet{Lorenz1963}
\begin{align}\label{eq:lorenz}
  \frac{dx}{dt} &= \sigma (y-x)\\
  \frac{dy}{dt} &= x(\rho -z) -y\\
  \frac{dz}{dt} &= xy - \beta z
\end{align}
which produces chaotic \tser. Figure~\ref{fig:lorenz-ex} shows a 3D plot of the system using the standard parameters $\sigma=10$, $\rho=20$, and $\beta = \frac{8}{3}$, which Lorenz used in his \citeyear{Lorenz1963} paper.

Four \tser\ are generated to diagnose \pecc's effectiveness for well-characterized chaotic systems: one from the \henon\ map and three for the Cartesian coordinates of the Lorenz strange attractor. 

While the idealized sampling scheme described in Section~\ref{subsec:sampling-limits} uses the natural oscillatory \tscal\ of a \tser, it can be difficult to identify a baseline oscillatory period for a chaotic \tser. This set of chaotic examples demonstrate two ways to approximate a relevant $\tnat$ \tscal.

If the \tscal\ to probe is unknown, a rough oscillatory \tscal\ can be determined by identifying the locations (in time) of the local maxima (or local minima) of the \tser, calculating the time elapsed between consecutive peaks, and taking the average of those values. This is called the ``approximated $\tnat$" method.

Alternatively, the \tscal\ at which the maximum \ccomplexity\ ($C$) occurs can be used for the ideal sampling. The use of the maximum \ccomplexity\ ensures that the classification of the overall behavior of the system is as accurate as possible, since poorly sampled regular \tser\ will tend toward the linear regime and poorly sampled chaotic \tser\ will tend toward the stochastic regime. In this method, the \edel\ corresponding to the peak value in the $\Cofl$ curve\ is determined. From that value, the \pattscal\ $\tpat$ can be calculated (via Equation~\ref{eq:tpat}). Depending on the $\tpat/\tnat$ ratio used (typically 0.4), that $\tpat$ value can be used to find the natural oscillatory \tscal\ $\tnat$. This is called the ``$\tnat$ from maximum $C(\edelsymbol)$" method.

\begin{figure*}
    \centering
    \includegraphics[width=\textwidth, trim=2cm 0cm 2cm 0cm]{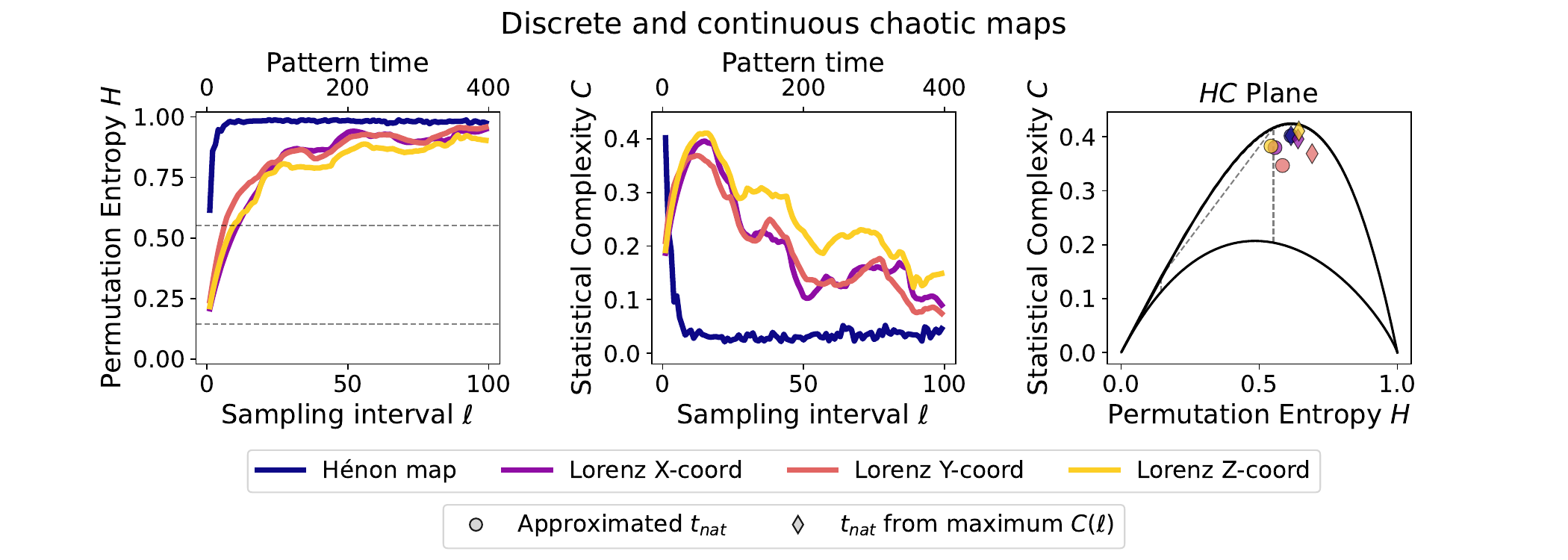}
    \caption{Values of \hentropy\ and \ccomplexity\ values for a range of \edel s for the discrete \henon\ map and the continuous $x$/$y$/$z$-coordinates of the Lorenz strange attractor. \textit{Left}: \hentropy\ as a function of \edel/\pattscal. For all \tser, the $H$-values increase initially and approach 1 rapidly (for a discrete chaotic system) or gradually (for a continuous chaotic system). Horizontal dashed lines represent the lower and upper limits of $\HMinPer(n=5)$ and $\HMaxPer(n=5)$, respectively. \textit{Center}: \ccomplexity\ as a function of \edel/\pattscal. For the discrete chaotic map, the $C$-values drop rapidly and bottom out close to zero. With the continuous chaotic map, the values of $C$ initially increase and then gradually decrease. \textit{Right}: the $\hcval$ values plotted on the \hcplane\ for the different chaotic systems span across the \hcplane\ at ideal sampling. $\hcval$ points calculated using the ``approximated $\tnat$" method are shown as circles, while those determined with the ``$\tnat$ from maximum $\Cofl$" method are represented as diamonds.  All values fall within the chaotic regime.
    }
    \label{fig:chaos-ex}
\end{figure*}

Figure~\ref{fig:chaos-ex} illustrates the difference in how discrete and continuous chaotic maps behave in $\Hofl$ and $\Cofl$ curves when applying \pecc\ to the four different \tser. While both types of chaotic maps increase in \hentropy\ $H$ as the \edel\ $\edelsymbol$ increases, in \ccomplexity, the discrete map falls from its initial value and stays constant, while the continuous map increases and then eventually drops as the \edel\ increases. The \hcplane\ shows the $\hcval$ values calculated from ideal sampling with the ``approximated $\tnat$" method as circles and those calculated with the ``$\tnat$ from maximum $\Cofl$" method as diamonds. All points fall within the chaotic regime of the \hcplane.

This exercise illustrates that a chaotic \tser\ may appear to be entirely stochastic if it is sampled at intervals where preferred or forbidden patterns cannot be resolved. A chaotic signal may have multiple characteristic \tscals\ for preferred and forbidden patterns, but these can only be discerned within the \tscals\ explored by the selected range of \edel. In addition, the optimal \edim, $n$, can be modified in order to sample patterns of varying length or complexity.

\subsection{Double Pendulum}\label{subsubsec:exs-dpen}
\begin{figure*}
    \centering
    \includegraphics[width=\textwidth, trim=2cm 0cm 2cm 0cm]{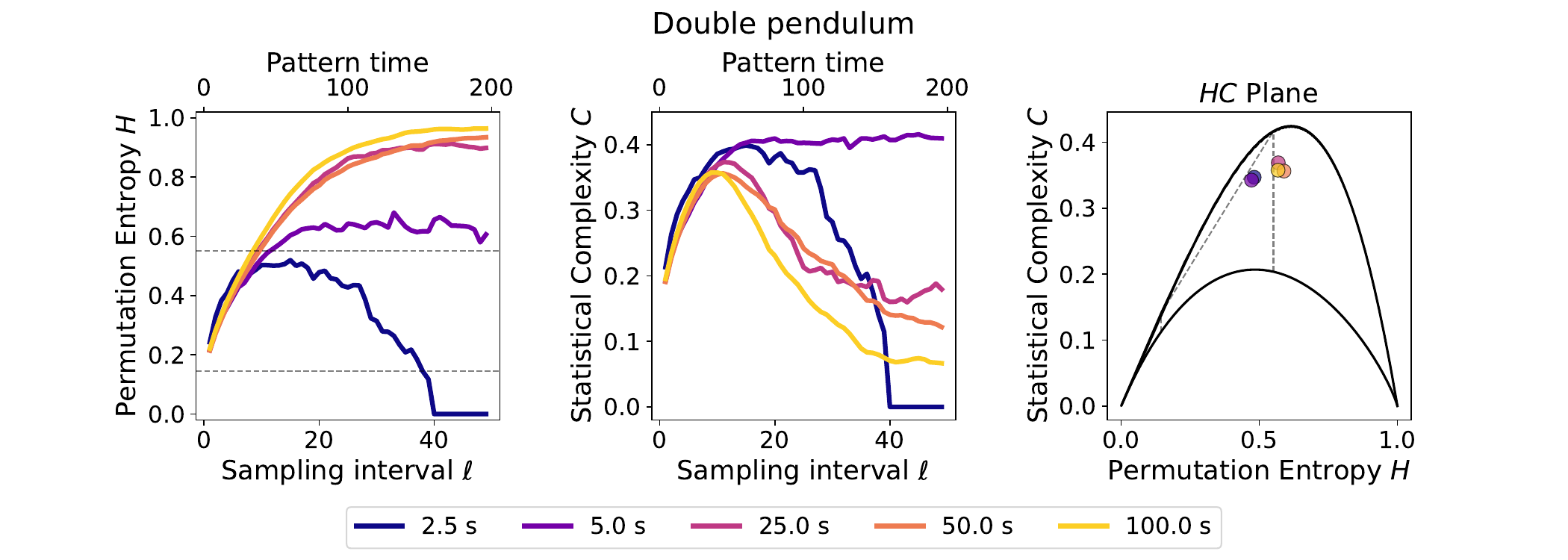}
    \caption{Values of \hentropy\ and \ccomplexity\ values for a range of \edel s for the double pendulum. \textit{Left}: \hentropy\ as a function of \edel/\pattscal. For all \tser, the $H$-values increase gradually. Horizontal dashed lines represent the lower and upper limits of $\HMinPer(n=5)$ and $\HMaxPer(n=5)$, respectively. \textit{Center}: \ccomplexity\ as a function of \edel/\pattscal. For all simulation \ durations, the values of $C$ initially increase and then gradually decrease, with the exception of the shortest simulations. \textit{Right}: the $\hcval$ values plotted on the \hcplane\ for the different simulation durations at ideal sampling. All values fall within the complex regime, with the exception of the 2.5~s simulation.}
    \label{fig:dpen-ex}
\end{figure*}

\pecc's \texttt{examples.doublePendulum} was used to generate \tser\ for a double pendulum system. The model assumes upper and lower pendulum masses of 1~kg each and pendulum lengths of 1~m. The system was allowed to evolve for a range of times (i.e., 2.5~s, 5~s, 10~s, 50~s, and 100~s) at a time resolution (i.e., step size) of $\delta t = 2^{-6}$~s. Figure~\ref{fig:dpen-ex} shows that while certain durations (e.g., 5~s and 10~s) remain in the complex region after turning off from the periodic/regular regime, too short a duration (e.g., 2.5~s) will appear regular. The sampling for the longer durations (50~s and 100~s) show that too long a sampling interval will cause chaotic behavior to appear as noise on the \hcplane. With the idealized sampling scheme, all the $\hcval$ points fall in the complex regime, with the exception of the too-short 2.5~s simulation.

\subsection{Astrophysical Examples}\label{subsec:exs-astro}

The tracer particle simulations for the astrophysical examples in the following subsections were created with \galpy\ \citep{Bovy15}, using the \texttt{symplec4\_c} integrator.

    \subsubsection{Keplerian Potential}\label{subsubsec:exs-kep}
\begin{figure}
        \centering
        \includegraphics[width=\columnwidth, trim=0cm 0cm 1.5cm 0.6cm]{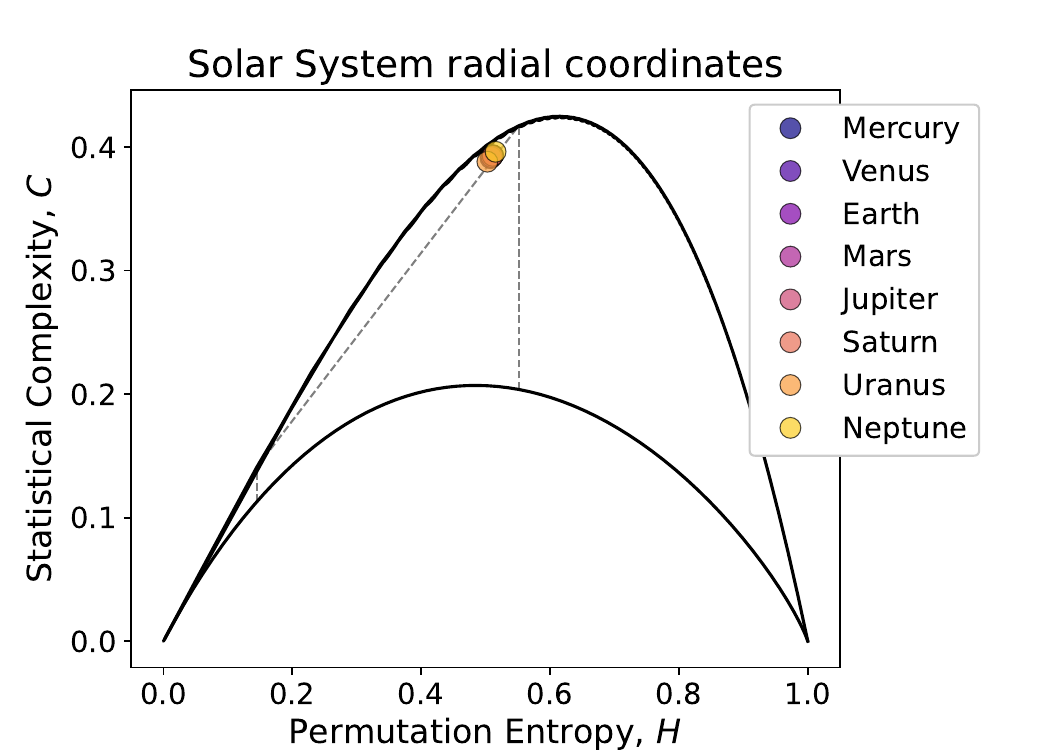}
        \caption{\hcplane\ showing the $\hcval$ values for tracer particle simulation radial coordinates of the solar System when $\tpattnat = 0.4$. The natural timescales ($\tnat$) are the radial periods of the orbits, determined with \galpy's \texttt{Orbit.Tr} function. All data points fall within the periodic/regular boundary, consistent with the fact that these are circular orbits.}
        \label{fig:kep-ex}
    \end{figure}

    While there are many types of regular orbits in astrophysics, only the point-mass (i.e., Keplerian) potential produces a special case of noncircular orbits that close in a single period, due to the fact that the radial and azimuthal frequencies are equal. As such, the Keplerian potential is an ideal scenario for testing regular orbits that close after $2\pi$. In this case, \galpy's \texttt{Orbit.from\_name(`solar system')} function was used to create a tracer particle simulation of the orbits of the eight planets of the solar system for a duration of 100~yr, at a resolution of 2.85~days. Figure~\ref{fig:kep-ex} demonstrates that all the values fall on and within the periodic/regular boundary of the \hcplane\ when using the radial coordinates for the orbits. The natural timescales used in the \pecc\ calculations for Figure~\ref{fig:kep-ex} are the radial periods of the planets' orbits, determined with \galpy's \texttt{Orbit.Tr} function.

    \subsubsection{Globular Cluster}\label{subsubsec:exs-gc}
\begin{figure}
        \centering
        \includegraphics[width=\columnwidth, trim=0cm 0cm 1.5cm 0.5cm]{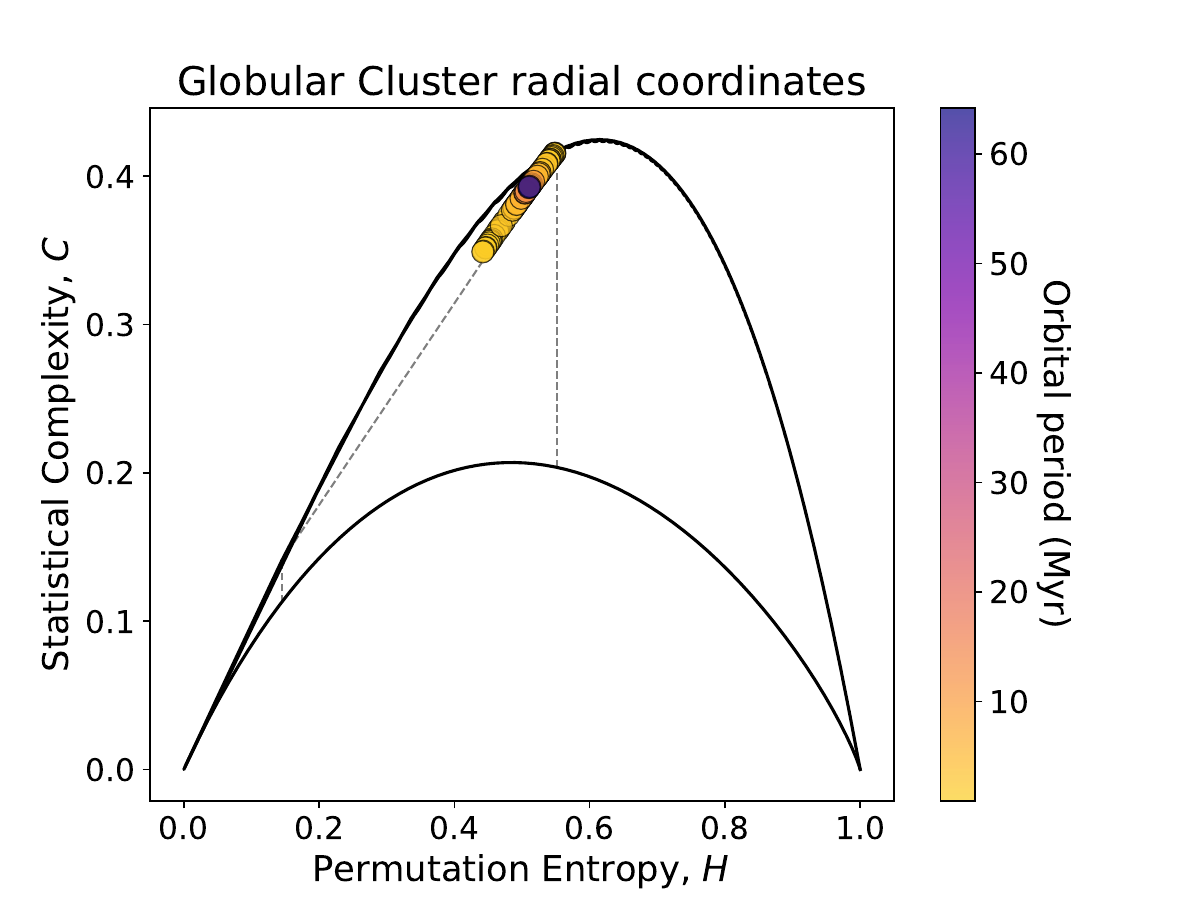}
        \caption{\hcplane\ showing the $\hcval$ values when $\tpattnat = 0.4$ for a sampling of 50 globular cluster tracer particles. The natural timescales ($\tnat$) are the radial periods of the orbits, determined with \galpy's \texttt{Orbit.Tr} function. The radial coordinates are used for the \pecc\ method, with orbital periods satisfying the $\tdurtnat \geq 1.5$ requirement. Most of the $\hcval$ values fall on or within the periodic/regular boundary/region, though some have lower complexity values. None of the $\hcval$ values for the orbits exceed the $\HMaxPer$ limit for $n=5$, consistent with the expectation that all of the orbits for a spherical potential are regular.}
        \label{fig:gc-ex}
    \end{figure}

    A spherical potential is a minimally intricate astrophysical example for testing the \pecc\ method. Using a spherical  Plummer potential (\texttt{galpy.potential.PlummerPotential}) and self-consistent isotropic and spherical Plummer distribution function (\texttt{galpy.df.isotropicPlummerdf}), $10^4$ tracer particles (representing stars) were evolved for a duration of 1~Gyr and an orbit integration time resolution of 0.1~Myr. Figure~\ref{fig:gc-ex} shows a sampling of 50 stellar orbits plotted on the \hcplane\ based on the $\hcval$ values calculated from the radial coordinates of orbits with $\tdur/\tpat \geq 1.5$ and $\tpattnat = 0.4$. The natural timescales ($\tnat$) used are the radial periods of the stellar orbits, determined with \galpy's \texttt{Orbit.Tr} function. As demonstrated in Figure~\ref{fig:gc-ex}, all orbits fall on the periodic/regular boundary of the \hcplane.
    
    \subsubsection{Triaxial Halo}\label{subsubsec:exs-triax}
    A triaxial potential will exhibit both regular and chaotic orbits. To verify that \pecc\ returns this same conclusion, a tracer particle simulation of a toy Navarro-Frenk-White (NFW) potential (\texttt{galpy.potential.TriaxialNFWPotential}), was created with an isotropic NFW distribution function that samples a spherical NFW halo (\texttt{galpy.df.isotropicNFWdf}). The triaxial NFW potential is defined by \galpy\ as 
    \begin{equation}
        \rho(x,y,z) = \frac{\mathcal{A}}{4 \pi a^3} \frac{1}{(m/a)(1+m/a)^{2}} \;,
    \end{equation}
    where $\mathcal{A}$ is the amplitude and $a$ is the scale radius. The parameter $m$ is described by
    \begin{equation}
        m^2 = x^2 + \frac{y^2}{b^2}+\frac{z^2}{c^2}\;,
    \end{equation}
    where $x$, $y$, and $z$ are the Cartesian coordinates.
    
    The parameters $b$ and $c$ are the $y$-axis/$x$-axis and $z$-axis/$x$-axis ratios of the density, respectively. The default \galpy\ settings were used for the triaxial prescription, such that the amplitude of the potential is $\mathcal{A} = 1$ and the scale length is set to $a=2$. The default values for the axis ratios were changed to $b=0.7$ and $c=0.5$, which fall within the range of mean values of Milky Way-like halos in the Illustris-Dark (dark-matter-only) simulation, as reported by \citet{2019MNRAS.484..476C}. $10^4$ tracer particles were evolved for 10,000 \tstep s in default \galpy\ natural time units. 
    
\begin{figure}
        \centering
        \includegraphics[width=\columnwidth, trim=0cm 0cm 1.5cm 0.5cm]{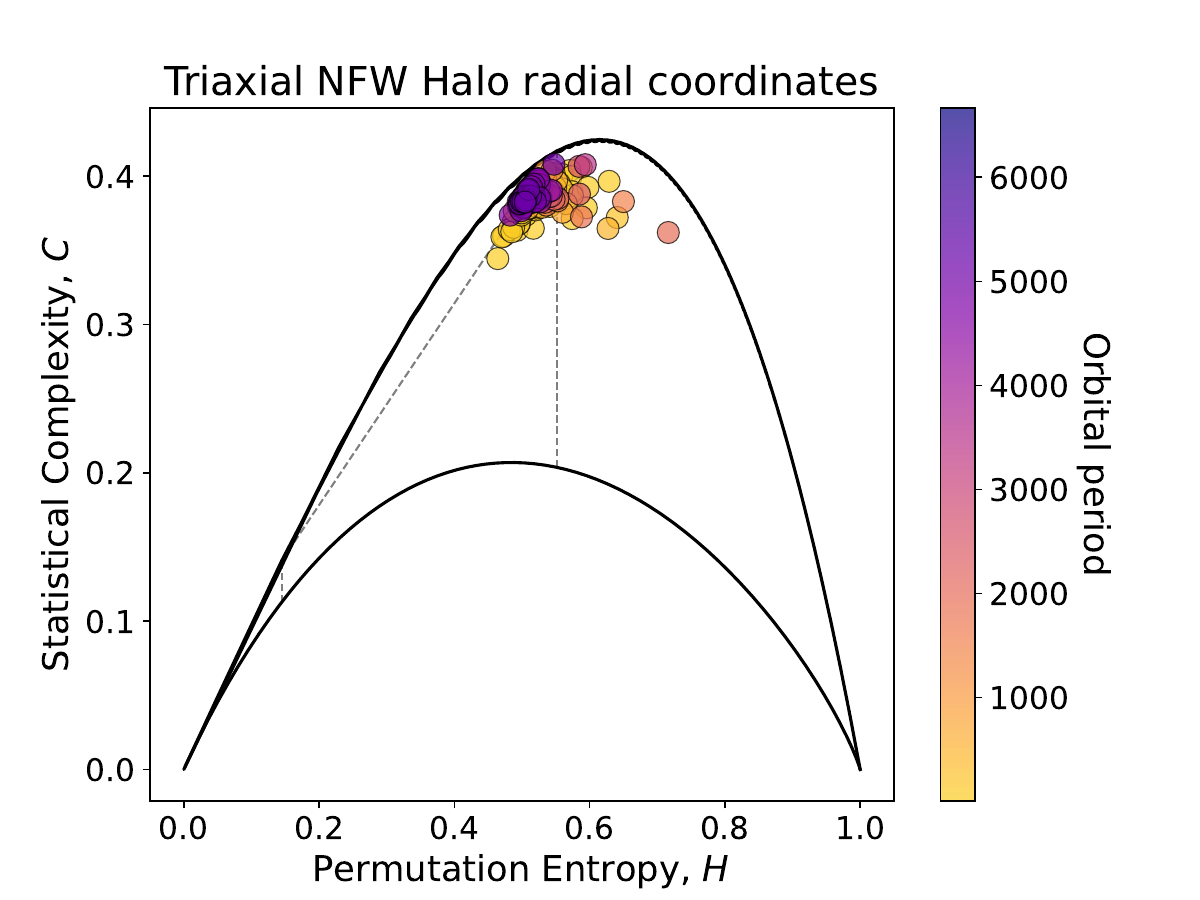}
        \caption{\hcplane\ showing the $\hcval$ values when $\tpattnat = 0.4$ for for a sampling of 1300 triaxial NFW halo tracer particles. The natural timescales ($\tnat$), which are given in \galpy\ natural units, are the approximated radial periods of the orbits, which are calculated by fitting a sine wave to the \tser. The radial coordinates are used for the \pecc\ method, with orbital periods satisfying the $\tdurtnat \geq 1.5$ requirement. As expected for a triaxial potential, there is a mix of particles falling in both the regular and complex zones.}
        \label{fig:triax-ex}
    \end{figure}

    Figure~\ref{fig:triax-ex} shows the $\hcval$ values at $\tpattnat = 0.4$ for the $r$-coordinates of a sampling of 1300 orbits with $\tdurtnat \geq 1.5$ (i.e., with radial oscillation periods up to 6666 \galpy\ \tstep s). Due to the non-axisymmetric nature of the potential, the \galpy\ \texttt{Orbit.Tr} calculations were not as reliable. Instead, the natural timescales ($\tnat$) were approximated by finding the best fit for a sine wave to each \tser. Orbits were classified as either regular or chaotic/complex, based on their $\hcval$ values. There are both regular and chaotic orbits present, with the number of regular orbits dominating. Future work will compare the results from \pecc\ to other known diagnostics of chaos.
    
\section{Future Work}\label{sec:future}
This paper introduces the \pecc\ method for usage in astrophysics. While the measures of \hentropy\ and \ccomplexity\ have been used in other fields, including plasma physics \citep{Weck2015}, to great success, those bodies of work involved systems or models that were inherently discrete and could use a \edel\ of $\edelsymbol=1$. The fundamentally continuous nature of many astrophysical systems, as well as the varied origin of stochasticity (i.e., natural noise or background sources/behaviors), requires great care in choosing the appropriate sampling schemes. The \pecc\ method provides the first clear recommendations for using \hentropy\ and \ccomplexity\ measures to characterize the behaviors of continuous systems.

This study investigates how well the method works for several astrophysical systems with known behaviors (Section~\ref{subsec:exs-astro}), but additional work is needed for widespread uses. Future papers will develop a method for estimating the confidence of the periodic/regular/chaotic/stochastic diagnosis and run comparisons with existing methods of chaos identification, such as frequency analysis mapping \citep{Laskar1990,ValluriMerritt98,Valluri12,Valluri16,BeraldoeSilva19,BeraldoESilva_2023} and potentially \Lyp\ exponents to test the robustness of \pecc\ in this regime.

Additional work is also needed to understand the sensitivities of the method for characterizing orbital behavior in a noisy signal for both idealized and realistic systems/simulations. Future research in this area will test how \pecc\ responds to injecting different levels of noise in frozen $n$-body simulations before expanding that to more intricate astrophysical simulations (e.g., evolving, time-dependent potentials and large-scale $n$-body simulations). Other work will explore the efficacy of stacking \tser\ in order to improve reliability for shorter-duration simulations and windowing, for understanding how a system changes dynamically over time.

\section{Conclusions}\label{sec:conclusion}
This paper introduces the \pecc\ method to the astrophysics community for the first time. \pecc\ is a statistical method that samples ordinal patterns from any sort of \tser, creates a probability distribution of all possible permutations of those patterns, and calculates the \hentropy\ $H$ and \ccomplexity\ $C$ from that distribution. The \tser\ is then classified as periodic, regular, complex, or stochastic based on its location in the \hcplane.

This paper provides an overview of the underlying theory and discusses best practices for initial implementations of the \pecc\ method. This work also demonstrates that for purely periodic functions, the orbital period can be easily extracted by using the shapes and initial peak of the $\Hofl$ curves.

The \pecc\ method is effective for \tser\ where the overall duration of the \tser\ is at minimum equal to the approximate period of the orbit, though a ratio of $\tdurtnat \gtrsim 1.5$ is ideal. While the overall shapes of the $\Hofl$ and $\Cofl$ curves provide the best indication for classifying the behavior of the data, in many cases it is better or more efficient to sample a single method. For cases such as these, the ratio between the \pattscal\ and the period should be between 0.3 and 0.5, i.e., $0.3 \lesssim \tpattnat \lesssim 0.5$. The corresponding \edel\ can be calculated with the \pecc\ package or using Equation~\ref{eq:tpat}.

Finally, a variety of different examples, both mathematical and astrophysical, are presented as a proof of concept of \pecc. Additional tests of the method's sensitivity, limitations, and wider applications will be presented in future papers in this series.

Extensive documentation and examples for the corresponding \pecc\ Python package \citep{peccaryZenodo} are available online.\footnote{\url{https://peccary.readthedocs.io}} The source code can be found on GitHub\footnote{\url{https://github.com/soleyhyman/peccary}} and builds can be found on the PyPI project page.\footnote{\url{https://pypi.org/project/peccary/}}

\begin{acknowledgments}
The authors thank Michael Petersen and Leandro Beraldo e Silva for their stimulating conversations and suggestions.
S.\'O.H. would like to acknowledge support from the University of Arizona's Theoretical Astrophysics Program's Travel Grant.  K.J.D. acknowledges support provided by the Heising-Simons Foundation grant \#2022-3927. 
The astrophysical simulations in Section~\ref{subsec:exs-astro} were created using High Performance Computing (HPC) resources supported by the University of Arizona TRIF, UITS, and Research, Innovation, and Impact (RII) and maintained by the UArizona Research Technologies department.
We respectfully acknowledge the University of Arizona is on the land and territories of Indigenous peoples. Today, Arizona is home to 22 federally recognized tribes, with Tucson being home to the O'odham and the Yaqui. The University strives to build sustainable relationships with sovereign Native Nations and Indigenous communities through education offerings, partnerships, and community service.
We recognize the Lenape Indian tribe as the original inhabitants of eastern Pennsylvania, where Bryn Mawr College stands. We acknowledge the Lenape people as the indigenous stewards of their homelands and also the spiritual keepers of the Lenape Sippu, or Delaware River. We respect and honor the ancestral caretakers of the land, from time immemorial until now, and into the future.

\end{acknowledgments}
\software{galpy \citep{Bovy15}, Astropy \citep{astropy:2013,astropy:2018,astropy:2022}, Numpy \citep{harris2020array_numpy}, Matplotlib \citep{Hunter:2007:Matplotlib}, SciPy \citep{2020SciPy-NMeth}}

\bibliography{references}{}
\bibliographystyle{aasjournalv7}



\end{document}